\documentclass[aps,twocolumn,pra,showpacs,longbibliography]{revtex4-1}

\usepackage{graphicx}
\usepackage{dcolumn}
\usepackage{bm}
\usepackage{amsmath}
\usepackage{amssymb}
\usepackage{amsthm}
\usepackage{amsfonts}
\usepackage{enumerate}
\usepackage{latexsym}
\usepackage[colorlinks=true,urlcolor=blue,citecolor=blue,linkcolor=blue]{hyperref}

\begin{document}
\title{Wave packet dynamics of Bogoliubov quasiparticles: quantum metric effects} 
\author{Long Liang}
\author{Sebastiano Peotta}
\author{Ari Harju}
\author{P\"{a}ivi T\"{o}rm\"{a}} \email{paivi.torma@aalto.fi}
\affiliation{COMP Centre of Excellence, Department of Applied Physics, Aalto University, Helsinki, Finland}
\begin{abstract}
We study the dynamics of the Bogoliubov  wave packet in superconductors and calculate the supercurrent carried by the wave packet. 
We discover an anomalous contribution to the supercurrent, related to the quantum metric of the Bloch wave function. This anomalous contribution is most important for flat or quasiflat bands, as exemplified by the attractive Hubbard models on the Creutz ladder and sawtooth lattice. Our theoretical framework is general and can be used to study a wide variety of phenomena, such as spin transport and exciton transport. 
\end{abstract}

\maketitle

\section{ introduction} 
 Charge transport in solids is one of the oldest problems in condensed matter physics. In the early days of the band theory of solids, the velocity of the Bloch electron was argued to be given by the group velocity, which is solely determined by the band dispersion \cite{solid_state_physics}. However, in the past several decades, it is increasingly clear that this description is incomplete. The Berry curvature \cite{Berry84}, a geometric property of the Bloch wave function, can drastically alter the transport properties \cite{PhysRev.95.1154,ADAMS1959286,PhysRevLett.62.2747,PhysRevLett.75.1348,SuNiu1999,PhysRevLett.93.206602,RevModPhys.82.1959,Wimmer2017},  
and it also plays an important role in the modern understanding of polarization  and orbital magnetization \cite{RevModPhys.66.899,PhysRevLett.95.137204,PhysRevLett.95.137205,resta2010}. 
The Berry curvature is the imaginary part of the quantum geometric tensor, whose real part gives another geometric quantity, the quantum metric \cite{Provost1980}, 
which measures the distance between Bloch states. Recently, the importance of the quantum metric is being revealed in  condensed matter physics 
  \cite{PhysRevLett.107.116801,PhysRevB.87.245103,roy2014,Roy2015,PT,LiebLattice,PhysRevB.95.024515,srivastava_signatures_2015,PhysRevLett.112.166601,GYN2015,PRFM2016,bleu_effective_2016,freimuth_geometrical_2017,2017arXiv170300712R}.

A simple yet powerful method to study the transport  of Bloch electrons is the semiclassical approximation. In this approach, the charge carriers are interpreted as wave packets sharply  localized in the momentum space. The evolution of the wave packet is described by the dynamics of its momentum and  center of mass,  
where the Berry curvature appears naturally \cite{SuNiu1999}. This formulation has been shown to be successful in a wide range of applications \cite{RevModPhys.82.1959}.    Very recently, it has been generalized to the second order of external electromagnetic field , and the quantum metric was  shown to play a 
	role in transport of the Bloch electrons only when the magnetic field is nonzero \cite{PhysRevLett.112.166601,GYN2015}.
 However, 
 this method has not been used in the study of  transport phenomena in superconductors, and  the Bogoliubov wave packet was explored only recently \cite{PhysRevLett.109.237004}. 
 
In this paper, we investigate the dynamics of the Bogoliubov  wave packet and analyze the supercurrent carried by it. Remarkably, we discover a geometric contribution to the supercurrent, which we call the {\it anomalous velocity}, in the sense that it {\it involves the quantum metric of the Bloch wave function and  does not depend on the group velocity of the Bloch electron}.  The integration of
the anomalous velocity gives rise to the geometric superfluid weight \cite{PT,LiebLattice,PhysRevB.95.024515}, which is  especially important for flat or quasiflat band superconductivity \cite{PhysRevB.83.220503,Volovik_Flatband2,FlatBandTc,PhysRevB.93.214505}.

To the best of our knowledge, this is the first work that identifies the anomalous velocity contribution to the supercurrent,
although  transport phenomena in superconductors have been intensively investigated using various methods, such as the Boltzmann equation \cite{Boltmann}, semiclassical  approximation based on physical arguments \cite{chandrasekhar_superconducting_1993} or   path integral formalism \cite{Vishwanath}, and more sophisticated quasiclassical  Green's function methods \cite{eilenberger_transformation_1968,larkin_quasiclassical_1968,rammer_quantum_1986,belzig_quasiclassical_1999,kopnin_theory_2001}.

 By using the
B\lowercase{ogoliubov-de }G\lowercase{ennes} (BdG)  Hamiltonian,  
   we go beyond the simplest $s$-wave pairing case \cite{PT,LiebLattice,PhysRevB.95.024515} and our results can be applied to the superconducting states with unconventional pairing symmetries \cite{RevModPhys.63.239}. 
Our theory is formulated for Bogoliubov quasiparticles; however, the essence of the results is rooted in the spinor structure of the wave function.
Therefore our theoretical framework is   general and can be applied to  a wide variety of  phenomena, such as  spin transport \cite{PhysRevB.68.241315,PhysRevLett.96.076604,PhysRevB.77.035327} and  exciton  \cite{keldysh1968collective} transport.

\section {Currents carried by  Bogoliubov quasiparticles} 
Our theoretical framework is general but for concreteness 
we focus on superconductors. We start from the BdG Hamiltonian, which captures the essential physics of superconducting states and  also describes other phenomena, such as exciton condensation \cite{keldysh1968collective} 
\begin{eqnarray}\label{Eq:BdG_Ham_Sec}
H&=&
\sum_{\sigma\sigma'}\int \mathrm{d}\mathbf{r}~c^\dag_\sigma(\mathbf{r}) [h_{\sigma\sigma'}(\mathbf{r})-\mu\delta_{\sigma\sigma'}] c_{\sigma'}(\mathbf{r})\nonumber\\
&&
+\int \mathrm{d}\mathbf{r} \mathrm{d}\mathbf{r}' ~[\Delta(\mathbf{r},\mathbf{r}')c^\dag_\uparrow(\mathbf{r})c^\dag_\downarrow(\mathbf{r}')+\mathrm{H.c.}],
\end{eqnarray}
where  $c^\dag_\sigma(\mathbf{r})$ is the operator that creates a free fermion with spin $\sigma=\uparrow, \downarrow$ at position $\mathbf{r}$, $\mu$ is the chemical potential, and $h_{\sigma\sigma'}(\mathbf{r})$ is the single particle Hamiltonian for fermions in a periodic potential.  For simplicity we assume that the single particle Hamiltonian preserves  the time reversal symmetry, which enables  us to write the BdG wave function in a simple way and therefore the geometric effects appear clearly. 
To simplify the notation, we take $h_{\sigma\sigma'}(\mathbf{r})=h_{\sigma}(\mathbf{r})\delta_{\sigma,\sigma'}$, and then  $h_{\uparrow}(\mathbf{r})=h^\ast_\downarrow(\mathbf{r})\equiv h(\mathbf{r})$, as a result of the time reversal symmetry. 
Furthermore, we focus on the spin singlet pairing potential $\Delta(\mathbf{r},\mathbf{r}')$, 
which  is assumed to be nonzero only if $\mathbf{r}-\mathbf{r}'$ is a lattice vector, and then it 
 can be factorized as 
$\Delta(\mathbf{r},\mathbf{r}')=\Delta_0(\mathbf{r})\delta_{\mathbf{x},\mathbf{x}'}\chi([\mathbf{r}]-[\mathbf{r}'])$, 
where $\mathbf{r}=[\mathbf{r}]+\mathbf{x}$ and 
$[\mathbf{r}]$ is the position of the unit cell and $\mathbf{x}$ is the position within the unit cell. 
This describes a large class  of possible pairings but not all.
The inter-unit cell part  $\chi([\mathbf{r}]-[\mathbf{r}'])$ determines the pairing symmetry, which is not necessarily an $s$-wave. 
The intra-unit cell part $\Delta_0(\mathbf{r})$ is a real and positive periodic function with the same periodicity as the periodic potential and can be understood as the modulus of the pairing potential. In the usual Bardeen-Cooper-Schrieffer (BCS) theory \cite{PhysRev.108.1175},  $\Delta_0(\mathbf{r})$ is approximated by a constant; however, it is generally position dependent  in the presence of periodic potential \cite{Tanaka1989491,PhysRevB.40.4482}.  
 We mention that our theory can also be generalized to include  spin orbit coupling and spin triplet pairing, see Appendix \ref{Appendix:1}.

 To study the supercurrent, we introduce a phase factor to the pairing potential, 
$\Delta(\mathbf{r},\mathbf{r}')\to e^{i \mathbf{q} \cdot(\mathbf{r+r}')}\Delta(\mathbf{r},\mathbf{r}')
.$ 
For convenience we will use the terminology ``electric current"; however,  our results can also be applied to a  charge neutral fermionic superfluid since the electric current we are studying is actually generated by the phase twist of the order parameter, and we do not require that the fermions carry true electric charge.

The supercurrent  can be obtained by evaluating the 
expectation value of the electric current operator
\begin{eqnarray}\label{Eq:j_operator}
\hat{\mathbf{j}}=\sum_{\sigma}\int \mathrm{d}\mathbf{r} ~c^{\dag}_\sigma(\mathbf{r})\hat{\mathbf{v}}_\sigma c_\sigma(\mathbf{r}),
\end{eqnarray}
where the single particle velocity operators are
$\hat{\mathbf{v}}_\uparrow=-\hat{\mathbf{v}}^\ast_\downarrow=-i [\hat{\mathbf{r}}_\uparrow,h(\mathbf{r})]$, with $\hat{\mathbf{r}}_\uparrow$ is the position operator of the up spin particle.
A crucial difference between a superconductor and a metal (or an insulator) is that, in a superconductor,  the electric current is  different from the quasiparticle current because a Bogoliubov quasiparticle is a mix of a particle and a hole and therefore its average charge is smaller than the charge of an electron \cite{Ronen16022016}. 
It is important to rewrite the current operator in terms of  Bogoliubov quasiparticles, which allows us to study the electric current carried by the Bogoliubov wave packet. To this end,  we turn to the more convenient BdG equation in Nambu form
\begin{eqnarray}\label{Eq:BdG_equation}
\int \mathrm{d}\mathbf{r}'\mathcal{H}_{\mathrm{BdG}}(\mathbf{r},\mathbf{r}')\psi_\alpha(\mathbf{r}')=\mathcal{E}_\alpha\psi_\alpha(\mathbf{r}),
\end{eqnarray}
with
\begin{eqnarray} 
 \mathcal{H}_{\mathrm{BdG}}(\mathbf{r},\mathbf{r}')=\left[\begin{array}{cc}
h(\mathbf{r})\delta_{\mathbf{r},\mathbf{r}'}&\Delta(\mathbf{r},\mathbf{r}')\\
\Delta^\ast(\mathbf{r}',\mathbf{r})& -h(\mathbf{r})\delta_{\mathbf{r},\mathbf{r}'}
\end{array}\right],
\end{eqnarray}
 the index $\alpha$ represents quantum numbers of the solutions, including  momentum and band index.  
The spinor $\psi_\alpha(\mathbf{r})=[u_\alpha(\mathbf{r}),v_\alpha(\mathbf{r})]^T$ is the wave function of the Bogoliubov quasiparticle, and  $u$ and $v$ are the particle and hole amplitudes, respectively.

Calculating the expectation value  of the electric current operator in the BCS state, we find (see Appendix \ref{Appendix:1} for details) %
\begin{eqnarray}\label{Eq:current_tot}
\mathbf{j}=-\frac{1}{2}\sum_{\alpha}\tanh\bigg(\frac{\beta\mathcal{E}_{\alpha}}{2}\bigg)\mathbf{j}_{e,\alpha}+\frac{1}{2}\sum_{\alpha}\mathbf{j}_{qp,\alpha},
\end{eqnarray}
where $\beta=1/(k_BT)$, $k_B$  the Boltzmann constant and $T$  the temperature,  $\mathbf{j}_{e,\alpha}=\langle \psi_{\alpha}| \hat{\mathbf{v}}_e |\psi_{\alpha}\rangle$ is  the quasiparticle charge current,  $\hat{\mathbf{v}}_e=\hat{\mathbf{v}}_\uparrow I$ and $I$ is the identity matrix in the particle-hole space, and $\mathbf{j}_{qp,\alpha}=\langle \psi_{\alpha}| \hat{\mathbf{v}}_{qp}|\psi_{\alpha}\rangle$ is the quasiparticle current, with $\hat{\mathbf{v}}_{qp}=-i[\hat{\mathbf{r}},H]$ and $\hat{\mathbf{r}}=\mathbf{r}I$    the velocity and position operators of the Bogoliubov quasiparticle, respectively. 
The existence of two types of currents in superconductors is known \cite{PhysRevB.25.4515} and the electric current has been separated into $\mathbf{j}_{qp,\alpha}$ and $\mathbf{j}_{e,\alpha}$ in the literature \cite{KopninSonin}. In this article we show that this separation is useful for the semiclassical approach. 
Intriguingly, we predict that $\mathbf{j}_{e,\alpha}$ can be finite even if $\mathbf{j}_{qp,\alpha}$ is zero, which means {\it there can still be electric current although the wave packet does not move}.

\section{wave packet dynamics and the supercurrent}
In general, the BdG Hamiltonian Eq.~\eqref{Eq:BdG_Ham_Sec} describes a multiband system. 
We here focus on the doubly-degenerate  Bloch bands that cross the Fermi level and assume that they are separated from other bands by sufficiently large gaps (isolated band approximation).  In the superconducting state, the Bloch bands become the Bogoliubov  bands, and we investigate the wave packet dynamics within these bands.

Within the isolated band approximation, the BdG equation, Eq.~\eqref{Eq:BdG_equation}, can be solved using the following ansatz
\begin{eqnarray}\label{Eq:qp_wavefunction}
\psi_{\mathbf{k}}(\mathbf{r})=e^{i\mathbf{k}\cdot \mathbf{r}}\left[\begin{array}{c}
u_\mathbf{k}e^{i \mathbf{q} \cdot\mathbf{r}}m_\mathbf{k+q}(\mathbf{r}) \\ 
v_\mathbf{k}e^{-i \mathbf{q} \cdot\mathbf{r}}m_\mathbf{k-q}(\mathbf{r})
\end{array} \right],
\end{eqnarray}
where $m_{\mathbf{k}}(\mathbf{r})$ is the periodic part of the Bloch function of the up spin band. The  Berry connection $a_{i}(\mathbf{k})$ and the quantum metric  $g_{ij}(\mathbf{\mathbf{k}})$ of the Bloch band are defined through $m_{\mathbf{k}}(\mathbf{r})$,
\begin{eqnarray}
a_{i}(\mathbf{k})&=&-i\langle m_\mathbf{k}  |\partial_{i}|m_\mathbf{k} \rangle,\label{Eq:a}\\
g_{ij}(\mathbf{k})&=&2\Re\langle \partial_{i} m_\mathbf{k}| (1-|m_\mathbf{k}\rangle\langle m_\mathbf{k}|)|\partial_{j}m_\mathbf{k} \rangle,\label{Eq:g}
\end{eqnarray}
where   $i,j=x,y,z$ are spatial indices, and $\partial_i\equiv \partial_{k_i}$ means  the derivative with respect to $k_i$.
The spinor $(u_{\mathbf{k}},v_{\mathbf{k}})^T$ is the Bogoliubov wave function in the Bloch basis. The physical picture behind this ansatz is clear: in the $\mathbf{q}=0$ limit, it describes a Cooper pair formed by Bloch electrons with opposite momentum and spin. For finite $\mathbf{q}$, the Cooper pair (with the wave function 
proportional to $e^{i2\mathbf{q}\cdot \mathbf{r}}u_\mathbf{k}v^\ast_\mathbf{k}$) acquires nonzero total momentum and therefore carries electric current.

The spinor $(u_{\mathbf{k}},v_{\mathbf{k}})^T$ can be determined by solving the eigenvalue problem (see Appendix \ref{Appendix:2_1}),
\begin{eqnarray}\label{Eq:BdG_matrix}
\left[\begin{array}{cc}
 \xi_\mathbf{k+q} & \Delta_{\mathbf{k}}(\mathbf{q}) \\ 
\Delta^\ast_{\mathbf{k}}(\mathbf{q}) & -\xi_\mathbf{k-q}
\end{array} 
\right]\left[ \begin{array}{c}
 u_\mathbf{k}\\ 
v_\mathbf{k}
\end{array} \right]=\mathcal{E}_\mathbf{k}(\mathbf{q})\left[ \begin{array}{c}
 u_\mathbf{k}\\ 
v_\mathbf{k}
\end{array} \right],
\end{eqnarray}
where $\xi_{\mathbf{k}}=\varepsilon_{\mathbf{k}}-\mu$, with $\varepsilon_{\mathbf{k}}$ is the Bloch energy, $\Delta_{\mathbf{k}}(\mathbf{q})=\Delta_{0,\mathbf{k}}(\mathbf{q})\chi_{\mathbf{k}}$,  $\Delta_{0,\mathbf{k}}(\mathbf{q})=\langle m_\mathbf{k+q}|\Delta_0(\mathbf{r})| m_\mathbf{k-q}\rangle$, and   $\chi_{\mathbf{k}}$ is the Fourier transform of $\chi([\mathbf{r}]-[\mathbf{r}'])$. 
The eigenvalues of Eq.~\eqref{Eq:BdG_matrix}  are
\begin{eqnarray}
\mathcal{E}^s_{\mathbf{k}}(\mathbf{q})=\xi^-_{\mathbf{k}}(\mathbf{q})+sE_\mathbf{k}(\mathbf{q}), \label{Eq:BdG_Excitations}
\end{eqnarray}
where  $E_\mathbf{k}(\mathbf{q})=
 \sqrt{\xi^+_{\mathbf{k}}(\mathbf{q})^2+|\Delta_\mathbf{k}(\mathbf{q})|^2}$, $\xi^\pm_{\mathbf{k}}(\mathbf{q})=(\xi_\mathbf{k+q}\pm\xi_\mathbf{k-q})/2$,  and $s=\pm 1$ labels the upper and lower Bogoliubov  bands. The corresponding wave functions satisfy $u^+_\mathbf{k}=(v^{-}_\mathbf{k})^\ast$ and $v^+_\mathbf{k}=-(u^{-}_\mathbf{k})^\ast$.

\subsection{Wave packet and its dynamics}
The wave packet can be constructed using the  quasiparticle wave functions \cite{SuNiu1999} 
\begin{eqnarray}
\Psi^s_{\mathbf{k}^s_c}(\mathbf{r})=\int \mathrm{d}\mathbf{k}~ W^s_\mathbf{k}\psi^s_{\mathbf{k}}(\mathbf{r}),\label{Eq:Wavepacket}
 \end{eqnarray}
where $W^s_\mathbf{k}$ is a normalized distribution sharply localized around the mean wave vector $\mathbf{k}^s_c$. 
The center of mass of the wave packet has the same form as in  nonsuperconducting systems \cite{SuNiu1999},
$\mathbf{r}^s_c= \langle\Psi^{s}_{\mathbf{k}^s_c}|\hat{\mathbf{r}}| \Psi^s_{\mathbf{k}^s_c}\rangle
=-\big[\partial_{\mathbf{k}^s_c}\arg{W}^s_{\mathbf{k}^s_c}+\mathbf{A}^s(\mathbf{k}^s_c)\big]. 
$
 Here $\mathbf{A}^s(\mathbf{k})$ is the Berry connection of the Bogoliubov quasiparticle, consisting of contributions from the noninteracting Bloch function and the  spinor $(u_{\mathbf{k}},v_{\mathbf{k}})^T$.
 In nonsuperconducting systems, the mass center coincides with the charge center \cite{SuNiu1999}. However, this is not true in superconductors, where 
the charge center of the wave packet is given by  $\mathbf{r}^s_e= \langle\Psi^{s}_{\mathbf{k}^s_c}|\hat{\mathbf{r}}\tau^3| \Psi^s_{\mathbf{k}^s_c}\rangle$, with $\tau^3$ being the third Pauli matrix in the particle-hole space.  The charge center can be written as a function of $\mathbf{r}^s_c$ and $\mathbf{k}^s_c$, see Appendix \ref{Appendix:2_2}. 
 In general, the mass center and charge center are different, which makes the problem nontrivial.

The dynamics of the wave packet can be obtained from the time-dependent variational principle \cite{SuNiu1999,TDVP}.  
The equations of motion for the Bogliubov quasiparticles  possess the same form as for the Bloch electrons in solids \cite{SuNiu1999}
\begin{eqnarray}
\dot{\mathbf{r}}^s_c&=&\partial_{\mathbf{k}^s_c} \mathcal{E}^s_{\mathbf{k^s_c}}(\mathbf{q})+\dot{\mathbf{k}}^s_c\times \bm{\Omega}^s(\mathbf{k}^s_c),\\
\dot{\mathbf{k}}^s_c&=&\partial_{\mathbf{r}^s_c} \mathcal{E}^s_{\mathbf{k^s_c}}=0, \label{Eq:EOM_r}
\end{eqnarray}
where $\mathcal{E}^s_{\mathbf{k^s_c}}(\mathbf{q})$ replaces the noninteracting dispersion and $\bm{\Omega}^s(\mathbf{k}^s_c)=\nabla \times \mathbf{A}^s(\mathbf{k}^s_c)$ is the Berry curvature of the Bogoliubov quasiparticle, which actually does not appear in our system because the momentum is conserved. For inhomogeneous systems, like cold atomic gases in a harmonic trap \cite{RevModPhys.80.1215}, the energy will also depend on $\mathbf{r}^s_c$, and therefore the momentum is no longer conserved, giving a Berry curvature correction to the equation of motion of the mass center.  This approach  has  been used to study the Bose-Einstein condensate with a vortex \cite{PhysRevLett.97.040401}, in which case the Berry curvature plays an important role \cite{ao_berrys_1993,kopnin_spectral_1995,volovik_three_1995,sonin_magnus_1997,PhysRevLett.97.040401}. 
In this paper we  focus on  homogeneous systems where $\mathbf{k}^s_c$ is conserved, and  therefore it can be  replaced by 
$\mathbf{k}$ without confusion.

The quasiparticle current is directly given by $\dot{\mathbf{r}}^s_c$,
\begin{eqnarray}
 \mathbf{j}^s_{qp,\mathbf{k}}(\mathbf{q})=\dot{\mathbf{r}}^s_c=\partial_{\mathbf{k}} \mathcal{E}^s_{\mathbf{k}}(\mathbf{q}),\label{Eq:qp_group_velocity}
\end{eqnarray}
which in the small $\mathbf{q}$ limit is
\begin{eqnarray}\label{Eq:j_q_s}
 j^s_{qp,\mathbf{k},i}=s\partial_i E_\mathbf{k}+\partial_i\partial_j \varepsilon_\mathbf{k} q_j,
\end{eqnarray}
where  $E_\mathbf{k}= E_\mathbf{k}(\mathbf{q=0})$. 
As expected, the  quasiparticle current  is the group velocity of the Bogoliubov quasiparticle \cite{PhysRevB.25.4515}. In the presence of the periodic potential,  $\sum_{s,\mathbf{k}}\mathbf{j}^s_{qp,\mathbf{k}}$ is zero because $\varepsilon_{\mathbf{k}}$ is a periodic function of $\mathbf{k}$, so only the first term in Eq.~\eqref{Eq:current_tot}, the quasiparticle charge current, contributes to the electric current. For continuum systems without periodic potentials, $\partial_i\partial_j\varepsilon_{\mathbf{k}}$ gives the inverse mass of the particle. 
 Then for $i=j$, $\sum_{\mathbf{k}}\partial_{i}\partial_{j}\varepsilon_{\mathbf{k}}$ diverges  and cancels the divergence from $\mathbf{j}^s_{e,\mathbf{k}}$, see Eq.~\eqref{Eq:ve} and Appendix \ref{Appendix:2_3}.

To find the quasiparticle charge current, we write the Heisenberg equation of the charge position operator  $\hat{\mathbf{r}}\tau^3$ (see Appendix  \ref{Appendix:2_3})
\begin{eqnarray}
\frac{\mathrm{d} \hat{\mathbf{r}}\tau^3}{\mathrm{d} t} 
=\hat{\mathbf{v}}_e-\frac{\mathrm{d}H_{\mathrm{p}}}{\mathrm{d}\mathbf{q}}.\label{Eq:EOM_re}
\end{eqnarray}
Here $H_{\mathrm{p}}$ is the pairing part of the BdG Hamiltonian.  The last term in the above equation comes from the  rotation in the particle-hole space, so it does not contribute to the translational charge transport. This is like the spin transport in  spin orbit coupled systems, where the spin current associated with the spin rotation does not contribute to the translational transport \cite{PhysRevB.77.035327}. 
Also, the velocity 
 $\mathrm{d}( \hat{\mathbf{r}}\tau^3)/\mathrm{d} t$ is analogous to the spin current defined in \cite{PhysRevLett.96.076604}. Because of these similarities, the theoretical framework developed here may also be used to study both the conventional \cite{PhysRevB.68.241315} and modified \cite{PhysRevLett.96.076604} spin currents. 

 From Eq.~\eqref{Eq:EOM_re} we see that the quasiparticle charge current is given by 
\begin{eqnarray}
\mathbf{j}^s_{e,\mathbf{k}}(\mathbf{q})=\dot{\mathbf{r}}^s_e+\langle \Psi^s_{\mathbf{k}}|\frac{\mathrm{d}H_{\mathrm{p}}}{\mathrm{d}\mathbf{q}}| \Psi^s_{\mathbf{k}}\rangle,
\end{eqnarray} 
furthermore, we find that (see Appendix \ref{Appendix:2_3})
  \begin{eqnarray}
   \mathbf{j}^s_{e,\mathbf{k}}(\mathbf{q})=\partial_{\mathbf{q}} \mathcal{E}^s_{\mathbf{k}}(\mathbf{q}). \label{Eq:j_e}
  \end{eqnarray}
 As we mentioned, $\mathbf{q}$ is the total momentum of a Cooper pair, so  $\partial_{\mathbf{q}}\mathcal{E}^s_{\mathbf{k}}(\mathbf{q})$ can be viewed as the group velocity of the Cooper pair, and therefore it gives the  charge current. Comparing Eq.~\eqref{Eq:qp_group_velocity} to Eq.~\eqref{Eq:j_e}, we conclude that $\mathcal{E}^s_{\mathbf{k}}(\mathbf{q})$ can be understood as the dispersion of both the quasiparticle and the Cooper pair, and  the quasiparticle  and charge currents are given by the group velocities of the quasiparticle and the Cooper pair, respectively.

Expanding Eq.~\eqref{Eq:j_e} to the first order of $\mathbf{q}$, we arrive at the most important result of this article, 
\begin{eqnarray}
&&\mathbf{j}^s_{e,\mathbf{k}}=\mathbf{v}^s_{e,\mathbf{k}}+\mathbf{v}^s_{a,\mathbf{k}},\label{Eq:js} 
\end{eqnarray}
with
\begin{eqnarray}
&&v^s_{e,\mathbf{k},i}=\partial_i \varepsilon_\mathbf{k}+s \frac{\xi_{\mathbf{k}}}{E_{\mathbf{k}}}\partial_i\partial_j \varepsilon_\mathbf{k} q_j,\label{Eq:ve}\\
&&
v^s_{a,\mathbf{k},i}=-2s\frac{|\Delta_{\mathbf{k}}|^2}{E_{\mathbf{k}}}\bar{g}_{ij}q_{j},\label{Eq:va}
\end{eqnarray}
where $\Delta_\mathbf{k}= \Delta_{\mathbf{k}}(\mathbf{q=0})$ is the order parameter without the phase twist, $\bar{g}_{ij}(\mathbf{k})$ is given by
\begin{eqnarray}
\bar{g}_{ij}(\mathbf{k})=g_{ij}(\mathbf{k})-\partial_{i}\partial_{j}\ln\Delta_0(\mathbf{k}),\label{Eq:gbar}
\end{eqnarray}
 with $\Delta_0(\mathbf{k})=\langle m_\mathbf{k}|\Delta_0(\mathbf{r})| m_\mathbf{k}\rangle$, and $g_{ij}(\mathbf{k})$ is the quantum metric of the modified Bloch function  $\tilde{m}_\mathbf{k}(\mathbf{r})=\sqrt{\Delta_0(\mathbf{r})/\Delta_0(\mathbf{k})}m_\mathbf{k}(\mathbf{r})$, which is defined by Eq.~\eqref{Eq:g}, with $m_{\mathbf{k}}(\mathbf{r})$ being replaced by $\tilde{m}_{\mathbf{k}}(\mathbf{r})$.

 The velocity $\mathbf{v}^s_{e,\mathbf{k}}$ may be understood in the following way:  the electric current  is 
carried by the particle and hole components of a Bogoliubov quasiparticle, so it may be written as $|u^s_{\mathbf{k}}|^2\mathbf{v}_p-|v^s_{\mathbf{k}}|^2\mathbf{v}_h$, where  
  $\mathbf{v}_p=\partial_{\mathbf{k}}\varepsilon_{\mathbf{k+q}}$ and $\mathbf{v}_h=-\partial_{\mathbf{k}}\varepsilon_{\mathbf{k-q}}$ are the group velocities of the particle and hole, respectively. Expanding $|u^s_{\mathbf{k}}|^2\mathbf{v}_p-|v^s_{\mathbf{k}}|^2\mathbf{v}_h$ to the first order of $\mathbf{q}$, we recover  Eq.~\eqref{Eq:ve}. Using a similar argument, the  superfluid weight (without the geometric contribution) was obtained in \cite{chandrasekhar_superconducting_1993}.  Here we show that this physical argument is partially validated by the systematic wave packet approach, and  most importantly, a new contribution, which is missing in this simple  argument, is revealed. 
  We call the newly discovered term, Eq.~\eqref{Eq:va}, the  anomalous velocity,
 in the sense that  it involves the geometric properties of the Bloch band and does not depend on the group velocity of the Bloch electron.
 
The anomalous velocity contributes to the superfluid weight (lattice equivalent of superfluid density) which tells whether the system is able to carry supercurrent. The anomalous velocity is of particular importance for flat or quasiflat bands where on the one hand critical temperatures are predicted to be greatly enhanced by the high density of states, but on the other hand the group velocity and conventional superfluid weight vanish. There the {\it geometric} part of the superfluid weight $D_{\mathrm{geom},ii}$ dominates. Using our results for the anomalous velocity we obtain from $D_{\mathrm{geom},ij}q_j=-\frac{1}{2}\sum_{s,\mathbf{k}}\tanh(\beta sE_{\mathbf{k}}/2)v^s_{a,\mathbf{k},i}$
 \begin{eqnarray}
D_{\mathrm{geom},ij}=2
\sum_{\mathbf{k}}\frac{|\Delta_{\mathbf{k}}|^2\tanh{(\beta E_\mathbf{k}/2)}}{E_\mathbf{k}}\bar{g}_{ij}(\mathbf{k}).\label{Eq:SW_g}
 \end{eqnarray}
 This is a generalization of previous results \cite{PT,LiebLattice,PhysRevB.95.024515} for superfluid weight, where the pairing potential  was restricted to be $\Delta(\mathbf{r},\mathbf{r}')=\Delta_0\delta_{\mathbf{r},\mathbf{r}'}$. 
 Our new result can be  applied to  superconducting states with unconventional pairing symmetries, and it will be important to revisit the magnetic penetration depth measurements \cite{bozovic_dependence_2016} and assess the importance of the geometric term in unconventional superconductors.

 \subsection{Comparison to the fully quantum mechanical derivation}
  Using the semiclassical wave packet approach we have shown that the charge current is given by the group velocity of the Cooper pair, Eq.~\eqref{Eq:j_e}. The quantum metric enters the result because the excitation $\mathcal{E}^s_{\mathbf{k}}(\mathbf{q})$ contains the order parameter, which we have found to be in the small $\mathbf{q}$ limit directly connected to the modified quantum metric $\bar{g}_{ij}$ (see Appendix \ref{Appendix:2_1})
   \begin{eqnarray}
   \Delta_{0,\mathbf{k}}(\mathbf{q})=\Delta_{0}(\mathbf{k})\exp{[-2i a_{i}(\mathbf{k})q_i-\bar{g}_{ij}(\mathbf{k})q_{i}q_{j}]}. 
   \end{eqnarray}
Here 
  $a_{i}(\mathbf{k})$ is the Berry connection  of the modified Bloch function,  defined by Eq.~\eqref{Eq:a} with $m_{\mathbf{k}}(\mathbf{r})$ being replaced by $\tilde{m}_{\mathbf{k}}(\mathbf{r})$, and $\bar{g}_{ij}(\mathbf{k})$ involves the quantum metric of the modified Bloch function, see Eq.~\eqref{Eq:gbar}. The anomalous velocity comes from the $q^2$ correction to the order parameter. If the pairing potential $\Delta_0(\mathbf{r})$ is uniform in the orbitals that
  compose the band we are interested in  \cite{PhysRevB.94.245149}, $\bar{g}_{ij}$ reduces to the quantum metric of the noninteracting Bloch band.
  
   Since $\mathcal{E}^s_{\mathbf{k}}(\mathbf{q})$ is the  energy corresponding to the wave function, Eq.~\eqref{Eq:qp_wavefunction}, 
  one may think that the  result of Eq.~\eqref{Eq:j_operator} can be obtained by evaluating  the current  $\mathbf{j}_{e}$ 
  using the wave function.
However,  direct calculations  show that the anomalous contribution to $\mathbf{j}_e$ is missing, see Appendix \ref{Appendix:2_4}. The reason is that the wave function within the isolated band approximation, Eq.~\eqref{Eq:qp_wavefunction}, is accurate only up to the zeroth order of the inverse band gap and the interband processes \cite{PhysRevB.95.024515}  are not taken into account. 
To get the correct result in the fully quantum mechanical approach, we need to solve the BdG equation {\it by including all the bands and take the isolated band limit after obtaining the current.} 
The physics behind this procedure is opaque. 
On the other hand, the (lowest order) multiband effects have been incorporated in the energy $\mathcal{E}^s_{\mathbf{k}}(\mathbf{q})$, because the first order correction to the energy is obtained using the zeroth order wave function. In the semiclassical approach the currents are expressed in terms of $\mathcal{E}^s_{\mathbf{k}}(\mathbf{q})$, and therefore the multiband effects appear naturally.

\section{flat band ferromagnetism}
The theoretical framework developed in this paper may also be used to study other transport phenomena than superfluidity. As an  example, the result for  flat band superconductivity can be  applied to  flat band ferromagnetism  \cite{Tasaki01041998}. The only difference is that the electric current is replaced by the spin current. For definiteness, we  consider the repulsive Hubbard model.
Within the mean-field approximation, the  Hubbard interaction can be decoupled in the spin channel as
\begin{eqnarray}
H_{\mathrm{int}}\approx \int\mathrm{d}\mathbf{r}~[M(\mathbf{r})c^\dag_{\uparrow}(\mathbf{r})c_{\downarrow}(\mathbf{r})+\mathrm{H.c.}],
\end{eqnarray}
with $M(\mathbf{r})=U\langle c^\dag_\downarrow(\mathbf{r}) c_\uparrow(\mathbf{r})\rangle$.
Assuming  $M(\mathbf{r})=M_0 e^{i\mathbf{Q}\cdot \mathbf{r}}$, then the single band mean-field Hamiltonian reads 
\begin{eqnarray}
&&H=\sum_\mathbf{k}\mathbf{c}^\dag_{\mathbf{k}}\left[
\begin{array}{cc}
\xi_\mathbf{k} & M_0 \langle m_\mathbf{k}|m_\mathbf{k+Q} \rangle \\
M_0 \langle m_\mathbf{k+Q}|m_\mathbf{k} \rangle  & \xi_\mathbf{k+Q}
\end{array}\right]\mathbf{c}_{\mathbf{k}},~~
\end{eqnarray}
where $\mathbf{c}_{\mathbf{k}}=(
c_{\mathbf{k}\uparrow},
c_{\mathbf{k+Q}\downarrow}
)^T$.  
In general, a finite interaction strength is required to trigger the ferromagnetic instability \cite{stoner}. However, for the flat band with  
$\xi_\mathbf{k}=0$,  there is magnetic instability for any nonzero repulsive interaction. 
 The ferromagnetic state with $\mathbf{Q=0}$ has the lowest energy because the overlap of the Bloch functions reaches the maximum.  Within this mean-field approximation of the flat band ferromagnetism,  the spin center is analogous to the charge center and it is immediately clear [cf. Eq.~\eqref{Eq:EOM_re}] that the spin current is given by the anomalous velocity, Eq.~\eqref{Eq:va}, with the pairing order parameter $\Delta_0$ being replaced by the magnetization $M_0$. Moreover, the superfluid weight Eq.~\eqref{Eq:SW_g} corresponds to the spin stiffness.

\section{Illustrative modes}
Having established the currents carried by Bogoliubov  wave packets, we now study two concrete models to confirm the validity of our theory and  to illustrate the effect of the anomalous velocity.
\subsection{The attractive Hubbard model on the Creutz ladder}
\begin{figure} [h]
\includegraphics[width=\columnwidth]{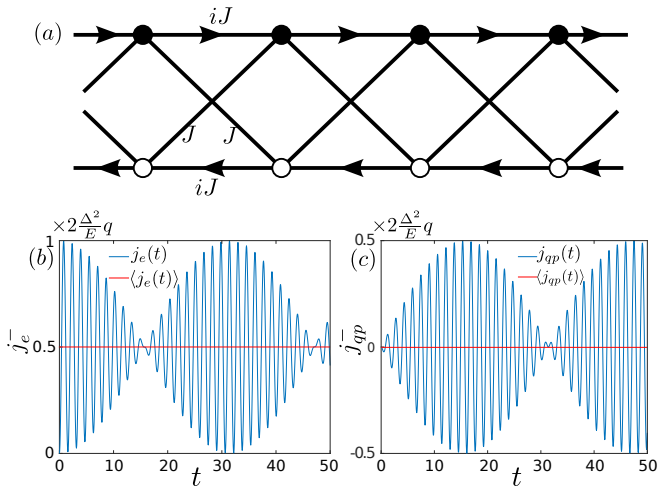}
\caption{(a), The Creutz ladder: the hopping coefficients of spin up fermions are given on corresponding links. The arrows show the directions of the positive phase for the complex nearest-neighbor hoppings. (b)-(c), charge and quasiparticle currents carried by the wave packet, obtained by simulating the motion of the wave packet. 
 The time  average of the currents agrees with our theory, $j^-_e = 2 g\frac{\Delta^2}{E}q$ and  $j^-_{qp} =0$. The quantum metric is a constant, $g=1/2$. }\label{Fig:j}
\end{figure}
We first study the attractive Hubbard model defined on  the Creutz ladder \cite{PhysRevLett.83.2636}, as shown in Fig.~\ref{Fig:j} (a).  In the noninteracting limit, it consists two perfectly flat bands with constant quantum metric $g=1/2$. For weak attractive Hubbard interactions, the BCS wave function is exact and the pairing potential $\Delta$ is uniform  \cite{PhysRevB.94.245149}. In principle $\Delta$ should be determined by solving the self-consistent equations. However, its value is not important here so we  treat it as a parameter.

To construct the wave packet with momentum $k_c$ and position $r_c$, we use the initial Gaussian distribution 
$W^\pm_k=\mathcal{N}e^{-(k-k_c)^2/4 k^2_0-i k r_c}$, 
where $\mathcal{N}$ is a normalization factor and $k_0$ is a parameter that controls the width of the wave packet in the momentum space. Because the quantum metric is a constant, the following results do not depend on $k_c$. 

The currents carried by the wave packet can be calculated as
$j^s_e(t)=\langle \Psi^s(t)|\hat{v}_e|\Psi^s(t)\rangle$ 
and $j^s_{qp}(t)=\langle \Psi^s(t)|\hat{v}_{qp}|\Psi^s(t)\rangle$,  
where $\Psi^s(t)=e^{-i H t}\Psi^s(t=0)$ is the time evolution of the wave packet. We calculate the currents for the lower band, and
the numerical results are shown in Figs.~\ref{Fig:j}~(b) and \ref{Fig:j}~(c). The currents oscillate in time, and their  time averages agree with our theory. Remarkably,  the wave packet can transport charge without  net displacement.

\subsection{The attractive Hubbard model on the sawtooth lattice}
\begin{figure} 
\includegraphics[width=\columnwidth]{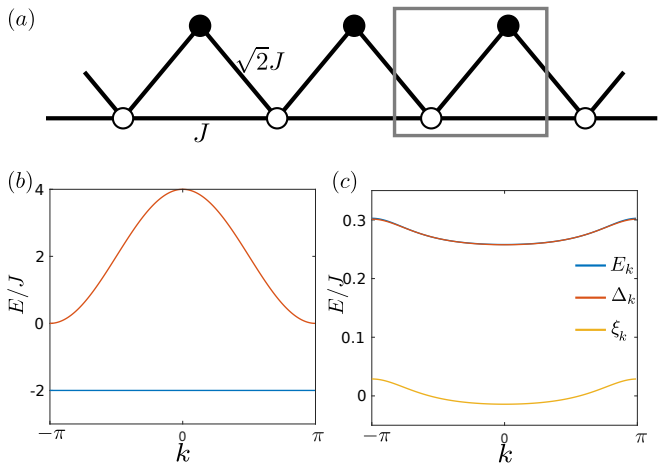}
\caption{(a), The sawtooth lattice and its unit cell (grey box). (b), dispersions of the noninteracting model. The lower band is flat. (c),  Bogoliubov dispersion for the flat band of the noninteracting limit. The interaction strength is $U/J=1$. The filling is chosen such that the flat band is half-filled in the noninteracting limit.  Within the isolated band approximation, $E_{k}=\sqrt{\xi^2_k+\Delta^2_{k}}$.}\label{Fig:Saw_Tooth_lattice}
\end{figure}

Now we consider another example, the attractive Hubbard model on the sawtooth lattice \cite{PhysRevB.81.014421,PhysRevB.82.184502}, sketched in Fig.~\ref{Fig:Saw_Tooth_lattice} (a). In this case there is only one flat band in the noninteracting limit, as shown in Fig.~\ref{Fig:Saw_Tooth_lattice} (b). Moreover, the noninteracting quantum metric becomes momentum dependent. 
 The two sublattices within a unit cell [black and white circle in Fig.~\ref{Fig:Saw_Tooth_lattice} (a)] are inequivalent. Therefore, after turning on the attractive Hubbard interaction $-U$, the pairing order parameter $\Delta(r)$ is nonuniform, and  the noninteracting Hamiltonian is  modified by the Hartree field, see Appendix \ref{Appendix:3}. 
As a result, 
the dispersion of the Bogoliubov quasiparticle, for the band that is flat in the noninteracting limit, becomes nonflat, as shown in Fig.~\ref{Fig:Saw_Tooth_lattice} (c). 
The Bogoliubov dispersion $E_k$ is obtained by solving the mean-field Hamiltonian self-consistently. The filling is chosen such that the flat band is half-filled in the noninteracting limit.  Within the isolated band approximation, $E_{k}=\sqrt{\xi^2_k+\Delta^2_{k}}$, where 
$\Delta_{k}=\langle m_k|\Delta(r)|m_k\rangle$, and
$\xi_k$ and $|m_k\rangle$ are the energy  and the Bloch wave function in the presence of the Hartree field. 

\begin{figure} [ht]
\includegraphics[width=0.9\columnwidth]{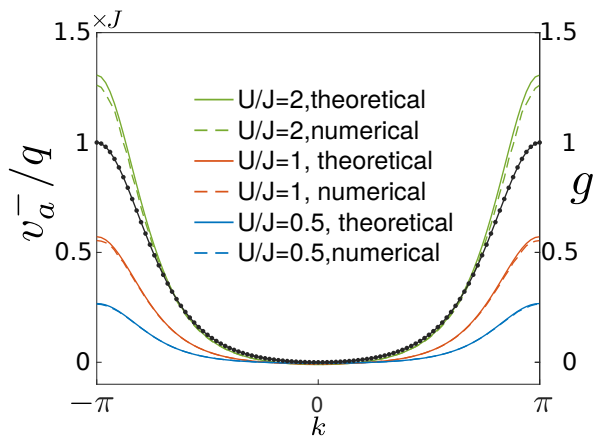}
\caption{The anomalous velocity for different interaction strengths at the same filling as in Fig.~\ref{Fig:Saw_Tooth_lattice} (c). Away from the Brillouin zone corner, the numerical results agree very well with our theory. The small deviation is because the band gap reaches the minimum at $k=\pi$. The agreement  becomes better with decreasing $U$.   The dotted black line is the  quantum metric of the noninteracting model, which has similar behavior as the anomalous velocity.
 }\label{Fig:Saw_Tooth}
\end{figure}

The time average of the quasiparticle and charge currents carried by the wave packet can be calculated using the method described in the previous section. To obtain the anomalous velocity, we first numerically calculate $j^-_{qp,k}$ and $j^-_{e,k}$ for both small and zero phase twists, and  separate the $q$ dependent current $\delta j_{qp/e,k}=j_{qp/e,k}(q)-j_{qp/e,k}(q=0)$. Then according to Eq.~\eqref{Eq:j_q_s} and Eqs.~\eqref{Eq:js}-\eqref{Eq:va}, the anomalous velocity can be extracted,
\begin{eqnarray}\label{Eq:va_n}
v^-_{a,k}=\frac{\xi_k}{E_k}\delta j^-_{qp,k}(q)+\delta j^-_{e,k}(q).
\end{eqnarray}

Fig.~\ref{Fig:Saw_Tooth} shows the anomalous velocities for various interaction strengths  at  the same filling as in Fig.~\ref{Fig:Saw_Tooth_lattice} (c), 
 calculated using Eq.~\eqref{Eq:va_n} (numerical results, solid lines) and Eq.~\eqref{Eq:va} (theoretical results, dashed lines). 
The numerical and theoretical results agree  well even at the corner of the Brillouin zone, where the band gap reaches the minimum and the isolated band approximation might not be good. As expected, the agreement is better with decreasing $U$. 
 The anomalous velocity and the noninteracting quantum metric have similar momentum dependencies, although the order parameter is nonuniform and the Bogoliubov dispersion is nonflat.

\section{conclusion}
We have analyzed the supercurrent carried by Bogoliubov quasiparticles. 
Using the powerful semiclassical wave packet approach, we discover a new contribution to the supercurrent, the anomalous velocity, which involves the quantum metric of the Bloch wave function. This contribution has been overlooked in previous literature. 
The  integration of the anomalous velocity gives rise to the geometric contribution of the superfluid weight. To validate
 our theory, we have studied two flat band models in which the effects of the anomalous velocity are clearly seen.

The magnetic penetration depth \cite{prozorov_magnetic_2006,prozorov_london_2011}, which is related to the superfluid weight, provides important information about the pairing states and  can be measured precisely \cite{bozovic_dependence_2016}. 
Our result of the superfluid weight can be  applied to  superconducting states with various pairing symmetries.
It is found that the superfluid weight in overdoped copper oxides is not given by the total electron density and this is interpreted as a failure of the BCS theory \cite{bozovic_dependence_2016}. However,  the usual BCS theory \cite{PhysRev.108.1175}
 neglects the effects of lattice, which are expected to be important in cuprates \cite{Zaanen2016}. 
 Our results show the intriguing possibility that taking into account the lattice effects (including the anomalous contribution) can explain features observed in high-$T_c$ superconductors.

The theoretical framework developed in this paper is  general and can be used to study also other phenomena than superfluidity. For example, 
because of the analogy between the electric current in superconductors and the spin current in non-superconducting systems, we predict that 
similar geometric effects also appear in  spin transport. The intriguing effects of Bloch wave functions in  condensed matter physics deserve  further study, and  the quantum metric may become a basic ingredient in our understanding of material properties.

\acknowledgments
We thank Grigory Volovik and Min-Fong Yang for useful comments. 
This work was supported by the Academy of Finland through its  Centres  of  Excellence  Programme  (2012-2017)  and under  Project  No.  263347,  No. 284621, and  No. 272490,  and by  the  European  Research  Council  (ERC-2013-AdG-340748-CODE). This project has received funding from the European Union's Horizon 2020 research and innovation programme under the Marie Sk\l{}odowska-Curie grant agreement No. 702281 (FLATOPS).

\appendix

\section{The B\lowercase{d}G Hamiltonian and the current operator}\label{Appendix:1}
Our  starting point is  the  BdG Hamiltonian
\begin{eqnarray}\label{Eq:BdG_Ham_Sec_G}
H&=&\sum_{\sigma\sigma'}\int \mathrm{d}\mathbf{r}~c^\dag_\sigma(\mathbf{r}) [h_{\sigma\sigma'}(\mathbf{r})-\mu\delta_{\sigma\sigma'}] c_{\sigma'}(\mathbf{r})\nonumber\\
&&
+\int \mathrm{d}\mathbf{r} \mathrm{d}\mathbf{r}' ~[\Delta_{\sigma\sigma'}(\mathbf{r},\mathbf{r}')c^\dag_\sigma(\mathbf{r})c^\dag_{\sigma'}(\mathbf{r}')+\mathrm{H.c.}].
\end{eqnarray}
We mainly focus on the cases where the noninteracting Hamiltonian  is time reversal invariant and block diagonal. Furthermore, the pairing potential is spin singlet and therefore is a scalar. Possible generalizations are  discussed at the end of this appendix.

In general, $h_\sigma(\mathbf{r})$ in the continuum form may be written as 
\begin{eqnarray}\label{Eq:h_a}
h_\sigma(\mathbf{r})=\frac{-\nabla^2_\sigma}{2}+V(\mathbf{r}),
\end{eqnarray}
where $V(\mathbf{r})$ is the periodic potential, $\nabla_\uparrow=\nabla+i \mathbf{A}(\mathbf{r})$, and $\nabla_\downarrow=\nabla-i \mathbf{A}(\mathbf{r})$, here $\mathbf{A}(\mathbf{r})$ is the vector potential that gives the periodic magnetic field whose periodicity is commensurate with the periodic potential. The mass, electric charge and Plank constant are taken to be unity. 
Our theory is formulated for the Hamiltonian in the continuum form; however, the results also apply to lattice models, which can be obtained from the continuum Hamiltonian through the tight-binding approximation.

We require that the pairing potential preserves the lattice translational symmetry, and then it can be written as
 \begin{eqnarray}
 \Delta(\mathbf{r},\mathbf{r}')=\Delta(\mathbf{x},\mathbf{x}',[\mathbf{r}]-[\mathbf{r}']),
 \end{eqnarray}
where  $\mathbf{r}=[\mathbf{r}]+\mathbf{x}$ and $[\mathbf{r}]$ is the position of the unit cell and $\mathbf{x}$ is the position within the unit cell. We assume that the inter-unit cell part and intra-unit cell part can be factorized, namely,
\begin{eqnarray}
 \Delta(\mathbf{r},\mathbf{r}')=\Delta_0(\mathbf{x},\mathbf{x}')\chi([\mathbf{r}]-[\mathbf{r}']).
\end{eqnarray}
 The pairing symmetry is determined by the inter-unit cell part $\chi([\mathbf{r}]-[\mathbf{r}'])$, which in general can be complex. For example, the  simplest isotropic $s$-wave pairing is 
 \begin{eqnarray}
 \chi([\mathbf{r}]-[\mathbf{r}'])=\delta_{[\mathbf{r}],[\mathbf{r}']}.
 \end{eqnarray}
 Assuming that the lattice has square symmetry, then
 the extended $s$-wave pairing is ($\mathbf{e}_x$ and $\mathbf{e}_y$ are primitive vectors)
 \begin{eqnarray}
 \chi([\mathbf{r}]-[\mathbf{r}'])=\sum_{s=\pm 1} (\delta_{[\mathbf{r}],[\mathbf{r}']+s \mathbf{e}_x}+\delta_{[\mathbf{r}],[\mathbf{r}']+s \mathbf{e}_y}),
 \end{eqnarray}
and the $d_{x^2-y^2}$-wave pairing is
\begin{eqnarray}
\chi([\mathbf{r}]-[\mathbf{r}'])=\sum_{s =\pm 1}(\delta_{[\mathbf{r}],[\mathbf{r}']+s\mathbf{ e}_x}-\delta_{\mathbf{[r]},[\mathbf{r}']+s\mathbf{ e}_y}).
\end{eqnarray}

We further assume that $\Delta_0(\mathbf{x},\mathbf{x}')=\Delta_0(\mathbf{x})\delta_{\mathbf{x},\mathbf{x}'}$, with $\Delta_0(\mathbf{x})$ is a real and positive function defined within a unit cell and  can be rewritten as a periodic function $\Delta_0(\mathbf{r})$. Physically, it  means that the pairing is nonzero only if 
the two electrons of a Cooper pair feel the same periodic potential (the distance between the two electrons is a multiple of the lattice vector),  
so this kind of pairing is likely the case for deep periodic potentials. 

The pairing potential
\begin{eqnarray}
 \Delta(\mathbf{r},\mathbf{r}')=\Delta_0(\mathbf{r})\delta_{\mathbf{x},\mathbf{x}'}\chi([\mathbf{r}]-[\mathbf{r}']),
\end{eqnarray}
 is not  the most general one, but it is already more general than the one usually used in the literature \cite{RevModPhys.63.239}. To see this,
  let us turn to the more familiar momentum space BdG  Hamiltonian.
Within the single band approximation,
we expand the  operator $c_{\sigma}(\mathbf{r})$ using the Bloch wave functions
\begin{eqnarray}
c_\sigma(\mathbf{r})=\sum_{\mathbf{k}} e^{i\mathbf{k}\cdot\mathbf{r}}m_{\mathbf{k}\sigma}(\mathbf{r})c_{\mathbf{k}\sigma},
\end{eqnarray}
here $c_{\mathbf{k}\sigma}$ annihilates a Bloch electron with momentum $\mathbf{k}$ and spin $\sigma$, $m$ is the band index denoting the band that crosses the Fermi level. The periodic part of the Bloch wave functions $m_{\mathbf{k}\sigma}(\mathbf{r})$ are related by the time reversal symmetry, $m_{\mathbf{k}\uparrow}(\mathbf{r})=m^\ast_{-\mathbf{k}\downarrow}(\mathbf{r})\equiv m_{\mathbf{k}}(\mathbf{r})$. Starting from Eq.~\eqref{Eq:BdG_Ham_Sec_G}, we obtain the widely used  phenomenological theory of superconductivity
\begin{eqnarray}\label{Eq:BdG_Ham_Sec_Bloch}
H&=&\int \mathrm{d}\mathbf{k}~\xi_{\mathbf{k}}c^\dag_{\mathbf{k}\sigma}c_{\mathbf{k}\sigma}+  [\Delta_0(\mathbf{k})\chi_{\mathbf{k}}c^\dag_{\mathbf{k}\uparrow}c^\dag_{-\mathbf{k}\downarrow}+\mathrm{H.c.}],~~~~
\end{eqnarray}
where $\xi_{\mathbf{k}}=\varepsilon_{\mathbf{k}}-\mu$, with $\varepsilon_{\mathbf{k}}$ is the Bloch energy, $\chi_{\mathbf{k}}$ is the Fourier transform of $\chi([\mathbf{r}]-[\mathbf{r}'])$ and
\begin{eqnarray}
\Delta_0(\mathbf{k})&=&\int_{\mathrm{u.c.}}\mathrm{d}\mathbf{r}\mathrm{d}\mathbf{r}'~ m^\ast_{\mathbf{k}}(\mathbf{r}')\Delta_0(\mathbf{r})\delta_{\mathbf{r},\mathbf{r}'} m_{\mathbf{k}}(\mathbf{r})e^{i\mathbf{k}\cdot(\mathbf{r}'-\mathbf{r})}, \nonumber\\
&=&\int_{\mathrm{u.c.}}\mathrm{d}\mathbf{r}~ m^\ast_{\mathbf{k}}(\mathbf{r})\Delta_0(\mathbf{r}) m_{\mathbf{k}}(\mathbf{r}),\nonumber\\
&=&\langle m_{\mathbf{k}}|\Delta_0(\mathbf{r}) |m_{\mathbf{k}}\rangle,
\end{eqnarray}
where  u.c. stands for unit cell. In previous literature \cite{RevModPhys.63.239},
the modulus of the pairing potential $\Delta_0(\mathbf{k})$ is usually approximated by a constant, indicating that $\Delta_0(\mathbf{r})$ is a constant.
In the presence of periodic potential, $\Delta_0$ is generally position dependent \cite{Tanaka1989491,PhysRevB.40.4482}, and our theory is able to capture this effect.

 Having  discussed the structure of the pairing potential and established the connection between the real space and momentum space BdG Hamiltonians, we turn to the  problem of the supercurrent.  As we will see, the Bloch wave function as well as the real space pairing potential $\Delta(\mathbf{r},\mathbf{r}')$ are needed to get the full supercurrent. 
We introduce a phase factor to the pairing potential, 
%
$\Delta(\mathbf{r},\mathbf{r}')\to e^{i \mathbf{q} \cdot(\mathbf{r+r}')}\Delta(\mathbf{r},\mathbf{r}')
$, to generate the supercurrent.

To study the dynamics of the Bogoliubov quasiparticle,  
it is convenient to work with the BdG equation in Nambu form, which can be viewed as the Schr{\"o}dinger equation for the Bogoliubov quasiparticle,
\begin{eqnarray}\label{Eq:BdG_equation_a}
\int \mathrm{d}\mathbf{r}'\mathcal{H}_{\mathrm{BdG}}(\mathbf{r},\mathbf{r}')\psi_\alpha(\mathbf{r}')=\mathcal{E}_\alpha\psi_\alpha(\mathbf{r}),
\end{eqnarray}
with
\begin{eqnarray} 
\mathcal{H}_{\mathrm{BdG}}(\mathbf{r},\mathbf{r}')=\left[\begin{array}{cc}
h_\uparrow(\mathbf{r})\delta_{\mathbf{r},\mathbf{r}'}&\Delta(\mathbf{r},\mathbf{r}')\\
\Delta^\ast(\mathbf{r}',\mathbf{r})& -h_\uparrow(\mathbf{r})\delta_{\mathbf{r}\mathbf{r}'}
\end{array}\right].\label{Eq:BdG}
\end{eqnarray}
The index $\alpha$ labels quantum numbers of the solutions, e.g., momentum and band index. The spinor $\psi_\alpha(\mathbf{r})=[u_\alpha(\mathbf{r}),v_\alpha(\mathbf{r})]^T$ is the wave function of the Bogoliubov quasiparticle, and  $u$ and $v$ are the particle and hole amplitudes, respectively. 

We define the position operator of the Bogoliubov quasiparticle  as $\hat{\mathbf{r}}=\mathbf{r}I$, where $I$ is the identity matrix in the particle-hole space. On the other hand, the charge operator of the Bogoliubov quasiparticle is given by the third Pauli matrix in the particle-hole space, $\tau^3$, and therefore the charge position operator of the Bogoliubov quasiparticle can be defined as $\hat{\mathbf{r}}\tau^3$.  

The  solution to the BdG equation $\psi_\alpha(\mathbf{r})=[u_\alpha(\mathbf{r}),v_\alpha(\mathbf{r})]^T$
 gives the Bogoliubov quasiparticle operator,
\begin{eqnarray}
  \gamma^\dag_\alpha=\int d\mathbf{r}~[ u_\alpha(\mathbf{r})c^\dag_{\uparrow}(\mathbf{r})+v_\alpha(\mathbf{r})c_{\downarrow}(\mathbf{r})],
\end{eqnarray}
 which diagonalizes the BdG Hamiltonian, 
 \begin{eqnarray}
 H=\sum_{\alpha}\mathcal{E}_{\alpha}\gamma^\dag_\alpha \gamma_\alpha.
 \end{eqnarray}
The operator $c_{\sigma}(\mathbf{r})$ can be written in terms of the Bogoliubov operators as
\begin{eqnarray}\label{Eq:c}
c_{\uparrow}(\mathbf{r})=\sum_{\alpha}u_{\alpha}(\mathbf{r})\gamma_{\alpha}, ~c^\dag_{\downarrow}(\mathbf{r})=\sum_{\alpha}v_{\alpha}(\mathbf{r})\gamma_{\alpha}.
\end{eqnarray} 
The electric current operator is
\begin{eqnarray}\label{Eq:j}
\hat{\mathbf{j}}=\sum_{\sigma}\int \mathrm{d}\mathbf{r} ~c^{\dag}_\sigma(\mathbf{r})\hat{\mathbf{v}}_\sigma c_\sigma(\mathbf{r}),
\end{eqnarray}
where the single particle velocity operator is
$\hat{\mathbf{v}}_\sigma=-i [\hat{\mathbf{r}}_\sigma,h_\sigma(\mathbf{r})]=-i\nabla_\sigma$ and $\hat{\mathbf{r}}_{\sigma}$ is the position operator of the  spin-$\sigma$ particle.  
Inserting  Eq.~\eqref{Eq:c} into  Eq.~\eqref{Eq:j} 
and evaluating its expectation value in the BCS state, 
we find the expression for the supercurrent 
\begin{eqnarray}
\mathbf{j}&=&\sum_{\alpha}\int \mathrm{d}\mathbf{r} ~f(\mathcal{E}_{\alpha})u^\ast_\alpha(\mathbf{r})(-i\nabla_\uparrow)u_\alpha(\mathbf{r})\nonumber\\
&&+\sum_{\alpha}\int \mathrm{d}\mathbf{r} ~[1-f(\mathcal{E}_{\alpha})]v_\alpha(\mathbf{r})(-i\nabla_\downarrow)v^\ast_\alpha(\mathbf{r}),\\
&=&\sum_{\alpha}\int \mathrm{d}\mathbf{r} ~f(\mathcal{E}_{\alpha})u^\ast_\alpha(\mathbf{r})(-i\nabla_\uparrow)u_\alpha(\mathbf{r})\nonumber\\
&&+\sum_{\alpha}\int \mathrm{d}\mathbf{r} ~[f(\mathcal{E}_{\alpha})-1]v^\ast_\alpha(\mathbf{r})(-i\nabla_\uparrow)v_\alpha(\mathbf{r}), 
\end{eqnarray}
where  $f(\mathcal{E}_{\alpha})$ is the Fermi-Dirac distribution.
We define the quasiparticle charge current $\mathbf{j}_{e,\alpha}$ and ``quasiparticle current'' $\mathbf{j}'_{qp,\alpha}$ as 
\begin{eqnarray}
&&\mathbf{j}_{e,\alpha} 
=\langle \psi_{\alpha}| \hat{\mathbf{v}}_e|\psi_{\alpha}\rangle= \int \mathrm{d}\mathbf{r}~\psi^\dag_\alpha(\mathbf{r})\hat{\mathbf{v}}_\uparrow I\psi_\alpha(\mathbf{r}),\label{Eq:je}\\
&&\mathbf{j}'_{qp,\alpha} 
=\langle \psi_{\alpha}| \hat{\mathbf{v}}_\uparrow\tau^3|\psi_{\alpha}\rangle=\int \mathrm{d}\mathbf{r}~\psi^\dag_\alpha(\mathbf{r})\hat{\mathbf{v}}_\uparrow\tau^3\psi_\alpha(\mathbf{r}),~~\label{Eq:j_qp}
\end{eqnarray}
Then the supercurrent can be written as
\begin{eqnarray}
\mathbf{j}
&=&-\frac{1}{2}\sum_{\alpha}\tanh\bigg(\frac{\beta\mathcal{E}_{\alpha}}{2}\bigg)\mathbf{j}_{e,\alpha}+\frac{1}{2}\sum_{\alpha}\mathbf{j}'_{qp,\alpha},\label{Eq:j_t_a}
\end{eqnarray}
where $\beta=1/(k_BT)$ with $k_B$ is the Boltzmann constant and $T$ is the temperature.

Although  $\mathbf{j}_{e,\alpha}$ appeared in  previous literature \cite{PhysRevB.25.4515,KopninSonin}, it does not have a name. Here we call it the quasiparticle charge current (or charge current for simplicity),  because it can be viewed as the electric current carried by the Bogoliubov quasiparticle.

The current $\mathbf{j}'_{qp,\alpha}$ needs more discussion.
For the isotropic $s$-wave paring, $\chi([\mathbf{r}],[\mathbf{r}'])=\delta_{[\mathbf{r}],[\mathbf{r}']}$, the pairing potential is {\it local} and commutes with the Bogoliubov position operator  $\hat{\mathbf{r}}$. Therefore $\hat{\mathbf{v}}_\uparrow\tau^3$ is the velocity operator of the quasiparticle,  $\hat{\mathbf{v}}_{qp}=-i[\hat{\mathbf{r}},H]=\hat{\mathbf{v}}_\uparrow\tau^3$. Then $\mathbf{j}'_{qp,\alpha}$ becomes the true quasiparticle current, $\mathbf{j}'_{qp,\alpha}=\mathbf{j}_{qp,\alpha} 
=\langle \psi_{\alpha}| \hat{\mathbf{v}}_{qp}|\psi_{\alpha}\rangle$, and we recover the result in \cite{PhysRevB.25.4515,KopninSonin}. However, for other pairing symmetries, the pairing potential becomes {\it nonlocal} and does not commute with the position operator. Therefore the quasiparticle velocity operator  contains an extra term, $-i[\hat{\mathbf{r}},H_\mathrm{p}]$, where $H_\mathrm{p}$ is the pairing part of the BdG Hamiltonian;  consequently, $\mathbf{j}'_{qp,\alpha}$  and $\mathbf{j}_{qp,\alpha}$  are different. However,  as we will see in Appendix \ref{Appendix:2_3}, $\sum_{\alpha}\langle \psi_{\alpha}|[\hat{\mathbf{r}},H_\mathrm{p}]|\psi_{\alpha}\rangle$ vanishes and therefore $\mathbf{j}'_{qp,\alpha}$ can be replaced by $\mathbf{j}_{qp,\alpha}$  in  Eq.~\eqref{Eq:j_t_a}, leaving the total current $\mathbf{j}$ unchanged. Finally, the supercurrent can be written as
\begin{eqnarray}
\mathbf{j}
&=&-\frac{1}{2}\sum_{\alpha}\tanh\bigg(\frac{\beta\mathcal{E}_{\alpha}}{2}\bigg)\mathbf{j}_{e,\alpha}+\frac{1}{2}\sum_{\alpha}\mathbf{j}_{qp,\alpha}.\label{Eq:j_t_a2}
\end{eqnarray}

Now we briefly discuss how to generalize our results to  systems with spin orbit coupling and spin triplet pairing. 
For spin orbit coupled systems, the noninteracting Hamiltonian is 
\begin{eqnarray}
h(\mathbf{r})&=&\frac{[-i\nabla+\mathbf{A}(\mathbf{r})\sigma^3]^2}{2}+V(\mathbf{r})
+\lambda[\nabla\cdot V_{\mathrm{so}}(\mathbf{r})\times \mathbf{p}]\cdot \bm{\sigma}, \nonumber\\ 
\end{eqnarray}
where $\sigma^i$ with $i=1,2,3$ is the Pauli matrix in the spin space and $\lambda$ denotes the strength of spin orbit coupling. 
 The BdG equation in the particle-hole space spanned by  $[c^\dag_\uparrow(\mathbf{r}),c^\dag_\downarrow(\mathbf{r}),-c_\downarrow(\mathbf{r}),c_\uparrow(\mathbf{r})]$  is
\begin{eqnarray}
 \mathcal{H}_{\mathrm{BdG}}(\mathbf{r},\mathbf{r}')=\left[\begin{array}{cc}
h(\mathbf{r})\delta_{\mathbf{r},\mathbf{r}'}&\Delta(\mathbf{r},\mathbf{r}')\\
\Delta^\ast(\mathbf{r}',\mathbf{r})&-h(\mathbf{r})\delta_{\mathbf{r},\mathbf{r}'}
\end{array}\right].
\end{eqnarray}
Here we have used the time reversal symmetry, i.e., $h(\mathbf{r})=i\sigma^2 h^\ast(\mathbf{r}) (-i\sigma^2)$. 
Similarly, the supercurrent is still given by Eq.~\eqref{Eq:j_t_a},
with the velocity operator $\hat{\mathbf{v}}_e$ given by
\begin{eqnarray}
\hat{\mathbf{v}}_e=\frac{1}{2}[-i\nabla+\mathbf{A}(\mathbf{r})\sigma^3+\lambda \bm{\sigma}\times \nabla\cdot V_{\mathrm{so}}(\mathbf{r})]I. 
\end{eqnarray}
 The  factor $1/2$ comes from the redundancy of the representation in the particle-hole space, i.e., both spin up and spin down operators appear in the particle and hole spaces. 

In the above derivations we do not require that the pairing is spin singlet and the expression for the supercurrent is unchanged for the  spin triplet pairing, although in general the pairing potential $\Delta(\mathbf{r},\mathbf{r}')$ becomes a matrix \cite{RevModPhys.63.239}.  
Therefore our semiclassical approach can be extended to the  general form of the BdG Hamiltonian, Eq.~\eqref{Eq:BdG_Ham_Sec_G}, without difficulty.


\section{Dynamics of the Bogoliubov wave packet}\label{Appendix:2}
In general, the Hamiltonian Eq.~\eqref{Eq:BdG_Ham_Sec_G} describes a  multiband system. We here focus on 
the mostly relevant bands, i.e., the  Bloch bands that cross the Fermi level. Because of the time reversal symmetry, they are doubly-degenerate. We further  assume that they are separated from other bands by sufficiently large gaps. This is the isolated band approximation \cite{PT,PhysRevB.95.024515}.
  In the superconducting state, the Bloch bands become the Bogoliubov  bands, and we investigate the  wave packet  dynamics within these bands.
 
  In this appendix we first solve the BdG Hamiltonian within the isolated band approximation. Using the solutions, we construct the Bogoliubov wave packet and
study its dynamics. The equations of motion of the  momentum and mass center of the wave packet are obtained, from which the equation of motion of the charge center can be derived. We then elaborate  the quasiparticle and charge currents carried by the wave packet and find that they are given by the group velocities of the qausiparticle and Cooper pair, respectively. The anomalous velocity, related to the quantum metric, appears naturally.
Finally, we compare to  the fully quantum mechanical derivation of the currents. We find that in the fully quantum mechanical approach, the isolated band wave function is not enough to obtain the correct results.

\subsection{Solutions to the BdG equation}\label{Appendix:2_1}
We first solve the BdG equation Eq.~\eqref{Eq:BdG_equation_a} within the isolated band approximation by using the ansatz 
\begin{eqnarray}\label{Eq:qp_wavefunction_a}
\psi_{\mathbf{k}}(\mathbf{r})=e^{i\mathbf{k}\cdot \mathbf{r}}\left[\begin{array}{c}
u_\mathbf{k}e^{i \mathbf{q} \cdot\mathbf{r}}m_\mathbf{k+q}(\mathbf{r}) \\ 
v_\mathbf{k}e^{-i \mathbf{q} \cdot\mathbf{r}}m_\mathbf{k-q}(\mathbf{r})
\end{array} \right],
\end{eqnarray}
where $m_{\mathbf{k}}(\mathbf{r})$ is the periodic part of the Bloch function of the spin up band we are interested in, with  Bloch energy $\varepsilon_{\mathbf{k}}$. 
The spinor $(u_{\mathbf{k}},v_{\mathbf{k}})^T$ is the Bogoliubov wave function in the Bloch basis. The physical picture behind this ansatz is clear: the wave function of a Cooper pair is
proportional to $e^{i2\mathbf{q}\cdot \mathbf{r}}u_\mathbf{k}v^\ast_\mathbf{k}$, so
in the $\mathbf{q}=0$ limit, it describes a Cooper pair formed by Bloch electrons with opposite momentum and spin, while for finite $\mathbf{q}$, the Cooper pair  acquires nonzero total momentum and therefore carries electric current.

 Substituting Eq.~\eqref{Eq:qp_wavefunction_a} into Eq.~\eqref{Eq:BdG_equation_a}, we get
\begin{widetext}
\begin{eqnarray}
&&\xi_\mathbf{k+q}e^{i(\mathbf{k+q})\cdot \mathbf{r}} m_\mathbf{k+q}(\mathbf{r})u_\mathbf{k}+\Delta_0(\mathbf{r}) \chi_{\mathbf{k}}e^{i(\mathbf{k+q})\cdot \mathbf{r}}m_\mathbf{k-q}(\mathbf{r})v_\mathbf{k}=\mathcal{E}_\mathbf{k}(\mathbf{q})e^{i(\mathbf{k+q})\cdot \mathbf{r}}m_\mathbf{k+q}(\mathbf{r})u_\mathbf{k},\label{Eq:B1}\\
&&\Delta_0(\mathbf{r})\chi^\ast_{\mathbf{k}}e^{i(\mathbf{k-q})\cdot \mathbf{r}}m_\mathbf{k+q}(\mathbf{r})u_\mathbf{k}-\xi_\mathbf{k-q}e^{i(\mathbf{k-q})\cdot \mathbf{r}} m_\mathbf{k-q}(\mathbf{r})v_\mathbf{k}=\mathcal{E}_\mathbf{k}(\mathbf{q})e^{i(\mathbf{k-q})\cdot \mathbf{r}}m_\mathbf{k-q}(\mathbf{r})v_\mathbf{k}.\label{Eq:B2}
\end{eqnarray}
\end{widetext}
As mentioned before, $\xi_\mathbf{k}=\varepsilon_{\mathbf{k}}-\mu$ and $\chi_\mathbf{k}$ is the Fourier transform of $\chi([\mathbf{r}]-[\mathbf{r}'])$. 
Projecting Eqs.~\eqref{Eq:B1} and~\eqref{Eq:B2} to the Bloch wave functions $e^{i(\mathbf{k+q})\cdot \mathbf{r}} m_\mathbf{k+q}(\mathbf{r})$ and $e^{i(\mathbf{k-q})\cdot \mathbf{r}} m_\mathbf{k-q}(\mathbf{r})$, respectively, we obtain the following eigenvalue problem
\begin{eqnarray} \label{Eq:BdG_matrix_a}
\left[\begin{array}{cc}
 \xi_\mathbf{k+q} & \Delta_{\mathbf{k}}(\mathbf{q}) \\ 
\Delta^\ast_{\mathbf{k}}(\mathbf{q}) & -\xi_\mathbf{k-q}
\end{array} 
\right]\left[ \begin{array}{c}
 u_\mathbf{k}\\ 
v_\mathbf{k}
\end{array} \right]=\mathcal{E}_\mathbf{k}(\mathbf{q})\left[ \begin{array}{c}
 u_\mathbf{k}\\ 
v_\mathbf{k}
\end{array} \right],~~~
\end{eqnarray}
where the momentum space order parameter in the presence of phase twist becomes $\Delta_{\mathbf{k}}(\mathbf{q})=\Delta_{0,\mathbf{k}}(\mathbf{q})\chi_{\mathbf{k}}$ with  $\Delta_{0,\mathbf{k}}(\mathbf{q})=\langle m_\mathbf{k+q}|\Delta_0(\mathbf{r})| m_\mathbf{k-q}\rangle$. 
 The eigenvalues and eigenvectors are obtained easily, 
%
\begin{eqnarray}
\mathcal{E}^s_{\mathbf{k}}(\mathbf{q})=\xi^-_{\mathbf{k}}(\mathbf{q})+sE_\mathbf{k}(\mathbf{q}),\label{Eq:BdG_Excitations_a}
\end{eqnarray}
where  $E_\mathbf{k}(\mathbf{q})=
\sqrt{\xi^+_{\mathbf{k}}(\mathbf{q})^2+|\Delta_\mathbf{k}(\mathbf{q})|^2}$, $\xi^\pm_{\mathbf{k}}(\mathbf{q})=(\xi_\mathbf{k+q}\pm\xi_\mathbf{k-q})/2$,  and $s=\pm 1$ labels the upper and lower Bogoliubov  bands. The corresponding wave functions can be chosen as
\begin{eqnarray}
&&u^+_\mathbf{k}=(v^{-}_\mathbf{k})^\ast=\frac{e^{i \arg(\Delta_\mathbf{k}(\mathbf{q}))/2}}{\sqrt{2}}\sqrt{1+\frac{\xi^+_\mathbf{k}(\mathbf{q})}{E_\mathbf{k}(\mathbf{q})}},\label{Eq:BdG_Wave1}\\
&&v^+_\mathbf{k}=-(u^{-}_\mathbf{k})^\ast=\frac{e^{-i\arg(\Delta_\mathbf{k}(\mathbf{q}))/2}}{\sqrt{2}}\sqrt{1-\frac{\xi^+_\mathbf{k}(\mathbf{q})}{E_\mathbf{k}(\mathbf{q})}}.\label{Eq:BdG_Wave2}
\end{eqnarray}

It is useful to expand $\Delta_{0,\mathbf{k}}(\mathbf{q})$ in the small $\mathbf{q}$ limit.
For convenience we define the modified Bloch function
\begin{eqnarray}
\tilde{m}_\mathbf{k}(\mathbf{r})=\sqrt{\Delta_0(\mathbf{r})/\Delta_0(\mathbf{k})}m_\mathbf{k}(\mathbf{r}),
\end{eqnarray}   
 with $\Delta_0(\mathbf{k})=\langle m_{\mathbf{k}}|\Delta_0(\mathbf{r}) |m_{\mathbf{k}}\rangle$ is positive. It is easily checked that   $\langle\tilde{m}_\mathbf{k}|\tilde{m}_\mathbf{k} \rangle=1$.
 
With the help of $\tilde{m}_\mathbf{k}(\mathbf{r})$, the pairing potential $\Delta_{0,\mathbf{k}}(\mathbf{q})$ can be written as
 \begin{eqnarray}
 \Delta_{0,\mathbf{k}}(\mathbf{q})&=&\sqrt{\Delta_0(\mathbf{k+q})\Delta_0(\mathbf{k-q})}\langle \tilde{m}_\mathbf{k+q}| \tilde{m}_\mathbf{k-q}\rangle.~~
 \end{eqnarray}
In the small $\mathbf{q}$ limit,
\begin{widetext}
\begin{eqnarray}
\ln \langle \tilde{m}_\mathbf{k+q}| \tilde{m}_\mathbf{k-q}\rangle&=&\ln\langle \tilde{m}_\mathbf{k}+ \partial_i \tilde{m}_\mathbf{k}q_i +\frac{1}{2}\partial_i\partial_j \tilde{m}_\mathbf{k}q_i q_j| \tilde{m}_\mathbf{k}  - \partial_n \tilde{m}_\mathbf{k}q_n +\frac{1}{2}\partial_n\partial_l m_\mathbf{k}q_n q_l \rangle
+O(q^3),\\
&=&\ln\{1-2\langle \tilde{m}_\mathbf{k}|\partial_i \tilde{m}_\mathbf{k}\rangle q_i-2\langle \partial_i \tilde{m}_\mathbf{k}|\partial_j \tilde{m}_\mathbf{k}\rangle q_i q_j \}+O(q^3),\\
&=& -2\langle \tilde{m}_\mathbf{k}|\partial_i \tilde{m}_\mathbf{k}\rangle q_i - 2\langle \partial_{i} \tilde{m}_\mathbf{k}| (1-|\tilde{m}_\mathbf{k}\rangle\langle \tilde{m}_\mathbf{k}|)|\partial_{j}\tilde{m}_\mathbf{k} \rangle q_i q_{j}+O(q^3).\label{Eq:ovlap} 
\end{eqnarray}
\end{widetext}
where $i,j,n,l=x,y,z$ are spatial indices and $\partial_i\equiv \partial_{k_i}$ means  the derivative with respect to $k_i$. 
It is easy to check that the first term in Eq.~\eqref{Eq:ovlap} is imaginary and the second term is real.
Using the Berry connection 
 \begin{eqnarray}
 a_{i}(\mathbf{k})=-i\langle\tilde{m}_\mathbf{k}  |\partial_{i}|\tilde{m}_\mathbf{k} \rangle,
 \end{eqnarray}
 and the quantum metric
 \begin{eqnarray}
 g_{ij}(\mathbf{k})=2\Re\langle \partial_{i} \tilde{m}_\mathbf{k}| (1-|\tilde{m}_\mathbf{k}\rangle\langle \tilde{m}_\mathbf{k}|)|\partial_{j}\tilde{m}_\mathbf{k} \rangle,
 \end{eqnarray}
Eq.~\eqref{Eq:ovlap} can be written as 
 \begin{eqnarray}
 \ln \langle \tilde{m}_\mathbf{k+q}| \tilde{m}_\mathbf{k-q}\rangle=-2i  a_i(\mathbf{k}) q_i - g_{ij}(\mathbf{k})q_i q_j+O(q^3)\nonumber.\\
 \end{eqnarray}
Denoting $\bar{g}_{ij}(\mathbf{k})=g_{ij}(\mathbf{k})-\partial_i\partial_j \ln \Delta_0(\mathbf{k})$, we find
\begin{eqnarray}
\Delta_{0,\mathbf{k}}(\mathbf{q})=\Delta_{0}(\mathbf{k})e^{-2i a_{i}(\mathbf{k})q_i-\bar{g}_{ij}(\mathbf{k})q_{i}q_{j}}+O(q^3).~~~\label{Eq:Delta0}
\end{eqnarray}
The quantum metric enters the supercurrent through this term. For the spin triplet pairing potential, which is in general a matrix,  we can also define $\Delta_{0,\mathbf{k}}(\mathbf{q})$. However, there is no obvious geometric structure in the small $\mathbf{q}$ limit.

For a constant $\Delta_0(\mathbf{r})$, $\Delta_0(\mathbf{k})$ is also a constant, and $\bar{g}_{ij}(\mathbf{k})$ reduces to the quantum metric of the noninteracting Bloch function. However, it is worth mentioning that this is not a necessary condition. It is enough that the pairing potential $\Delta_0(\mathbf{r})$ is uniform in the orbitals that compose the band we are interested in \cite{PhysRevB.94.245149}.
 %

\subsection{Bogoliubov wave packet and its dynamics}\label{Appendix:2_2}
Following Sundaram and Niu \cite{SuNiu1999}, we construct the wave packet from the wave fuction $\psi_\mathbf{k}(\mathbf{r})$ as
\begin{eqnarray}
\Psi^s_{\mathbf{k}^s_c}(\mathbf{r})=\int \mathrm{d}\mathbf{k}~ W^s_\mathbf{k}\psi^s_{\mathbf{k}}(\mathbf{r}),\label{Eq:Wavepacket_a}
\end{eqnarray}
where $s=\pm 1$ denotes the upper and lower Bogoliubov bands and $W^s_\mathbf{k}$ is a normalized distribution which is sharply localized around the mean wave vector $\mathbf{k}^s_c$. Mathematically, 
\begin{eqnarray}
&&\int\mathrm{d}\mathbf{k}~\mathbf{k}|W^s_{\mathbf{k}}|^2=\mathbf{k}^s_c, \\
&&\int\mathrm{d}\mathbf{k}~f(\mathbf{k})|W^s_{\mathbf{k}}|^2=f(\mathbf{k}^s_c), 
\end{eqnarray}
where $f(\mathbf{k})$ is an arbitrary function of $\mathbf{k}$. We can choose the same initial distributions $W^-_\mathbf{k}=W^+_\mathbf{k}$, and then the initial momenta  $\mathbf{k}^-_c$ and $\mathbf{k}^+_c$ are the same. However, their time evolutions can be different. 

 After a straightforward  calculation we find that the  mass center has the same form as in a metal \cite{SuNiu1999} 
\begin{eqnarray}
\mathbf{r}^s_c&= &\int \mathrm{d}\mathbf{r}~\Psi^{s\dag}_{\mathbf{k}^s_c}(\mathbf{r})\mathbf{r} \Psi^{s}_{\mathbf{k}^s_c}(\mathbf{r}),\nonumber\\
&=&\int \mathrm{d}\mathbf{r}\mathrm{d}\mathbf{k}\mathrm{d}\mathbf{k}' ~W^s_\mathbf{k} W^{s\ast}_{\mathbf{k}'}\mathbf{r}e^{i\mathbf{(k-k') r}}\phi^{s\dag}_{\mathbf{k}'}\phi^s_{\mathbf{k}}, \nonumber\\
&=&-i\int \mathrm{d}\mathbf{r}\mathrm{d}\mathbf{k}\mathrm{d}\mathbf{k}' ~W^s_\mathbf{k} W^{s\ast}_{\mathbf{k}'}(\partial_{\mathbf{k}} e^{i\mathbf{(k-k') r}} )\phi^{s\dag}_{\mathbf{k}'}\phi^s_{\mathbf{k}}, \nonumber\\
&=& i\int \mathrm{d}\mathbf{k} \int_{\mathrm{u.c.}} \mathrm{d}\mathbf{r} ~W^{s\ast}_\mathbf{k} \phi^{s\dag}_\mathbf{k} \partial_{\mathbf{k}}(W^s_\mathbf{k} \phi^s_\mathbf{k}) , \nonumber\\
&=& i\int \mathrm{d}\mathbf{k}  ~W^{s\ast}_\mathbf{k}  \partial_{\mathbf{k}}W^s_\mathbf{k}   -\int \mathrm{d}\mathbf{k} ~|W^s_\mathbf{k}|^2 \mathbf{A}^s(\mathbf{k}), \nonumber\\
&=&-\big[\partial_{\mathbf{k}^s_c}\arg{W}^s_{\mathbf{k}^s_c}+\mathbf{A}^s(\mathbf{k}^s_c)\big], 
\end{eqnarray}
where $\phi^{s}_{\mathbf{k}}(\mathbf{r})=e^{-i\mathbf{k}\cdot\mathbf{r}}\psi^{s}_{\mathbf{k}}(\mathbf{r})$ is analogous to the periodic part of the Bloch function and $\mathbf{A}^s(\mathbf{k})$ is the Berry connection of the Bogoliubov quasiparticle,
\begin{eqnarray}
\mathbf{A}^s(\mathbf{k})&=&-i\int_{\mathrm{u.c.}} \mathrm{d}\mathbf{r}~ \phi^{s\dag}_\mathbf{k} \partial_{\mathbf{k}} \phi^s_\mathbf{k},\nonumber\\
&=&-i (u^{s\ast}_\mathbf{k}\partial_\mathbf{k} u^{s}_\mathbf{k}+v^{s\ast}_\mathbf{k}\partial_\mathbf{k} v^{s}_\mathbf{k})\nonumber\\
&&+|u^{s}_\mathbf{k}|^2\mathbf{a}^0(\mathbf{k+q})+|v^{s}_\mathbf{k}|^2\mathbf{a}^0(\mathbf{k-q}).
\end{eqnarray} 
 Here $\mathbf{a}^0(\mathbf{k})$ is the Berry connection of the noninteracting Bloch state 
and $-i (u^{s\ast}_\mathbf{k}\partial_\mathbf{k} u^{s}_\mathbf{k}+v^{s\ast}_\mathbf{k}\partial_\mathbf{k} v^{s}_\mathbf{k})$ is the Berry connection of the wave function in the Bloch basis and is determined by the phase of the order parameter.
Similarly, the charge center is given by
\begin{eqnarray}
\mathbf{r}^s_e&=& \int \mathrm{d}\mathbf{r}~\Psi^{s\dag}_{\mathbf{k}^s_c}(\mathbf{r})\mathbf{r}\tau^3 \Psi^{s}_{\mathbf{k}^s_c}(\mathbf{r}),\nonumber\\
&=&\int \mathrm{d}\mathbf{r}\mathrm{d}\mathbf{k}\mathrm{d}\mathbf{k}' ~W^s_\mathbf{k} W^{s\ast}_{\mathbf{k}'}\mathbf{r}e^{i\mathbf{(k-k') r}}\phi^{s\dag}_{\mathbf{k}'}\tau^3\phi^s_{\mathbf{k}}, \nonumber\\
&=&-i\int \mathrm{d}\mathbf{r}\mathrm{d}\mathbf{k}\mathrm{d}\mathbf{k}' ~W^s_\mathbf{k} W^{s\ast}_{\mathbf{k}'}(\partial_{\mathbf{k}} e^{i\mathbf{(k-k') r}} )\phi^{s\dag}_{\mathbf{k}'}\tau^3\phi^s_{\mathbf{k}}, \nonumber\\
&=& i\int \mathrm{d}\mathbf{k} \int_{\mathrm{u.c.}} \mathrm{d}\mathbf{r} ~W^{s\ast}_\mathbf{k} \phi^{s\dag}_\mathbf{k}\tau^3 \partial_{\mathbf{k}}(W^s_\mathbf{k} \phi^s_\mathbf{k}) , \nonumber\\
&=&(|u^s_{\mathbf{k}^s_c}|^2-|v^s_{\mathbf{k}_c}|^2) \mathbf{r}^s_c
+[1-(|u^s_{\mathbf{k}^s_c}|^2-|v^s_{\mathbf{k}^s_c}|^2)^2]\nonumber\\
~~~~&&\times \partial_{\mathbf{k}^s_c}[\arg(\Delta_\mathbf{k}(\mathbf{q}))/2+a^0_{i}(\mathbf{k}^s_c)q_{i}].
\end{eqnarray}

The dynamics of the wave packet can be obtained from the time-dependent variational principle \cite{SuNiu1999}, and the
 effective Lagrangian is
 \begin{eqnarray}
 L^s&=&\langle \Psi^s_{\mathbf{k}^s_c}|i \partial_t-H |\Psi^s_{\mathbf{k}^s_c}\rangle\\
 &=&\mathbf{k}^s_c \cdot \dot{\mathbf{r}}^s_c-\dot{\mathbf{k}}^s_c \cdot  \mathbf{A}^s(\mathbf{k}^s_c) - \mathcal{E}^s_{\mathbf{k}^s_c}(\mathbf{q}).
 \end{eqnarray}
Then the equations of motion can be obtained,
\begin{eqnarray}
&&\dot{\mathbf{r}}^s_c=\partial_{\mathbf{k}^s_c} \mathcal{E}^s_{\mathbf{k^s_c}}(\mathbf{q})+\dot{\mathbf{k}}^s_c\times \bm{\Omega}^s(\mathbf{k}^s_c),\\
&&\dot{\mathbf{k}}^s_c=\partial_{\mathbf{r}^s_c} \mathcal{E}^s_{\mathbf{k^s_c}}=0, \label{Eq:EOM_r_a}
\end{eqnarray}
where $\bm{\Omega}^s(\mathbf{k}^s_c)=\nabla \times \mathbf{A}^s(\mathbf{k}^s_c)$ is the Berry curvature of the Bogoliubov quasiparticle. The center of mass $\mathbf{r}^s_c$ does not appear in the energy, so the momentum is conserved and the Berry curvature does not affect the equation of motion of $\mathbf{r}^s_c$. Since the momentum is conserved, $\mathbf{k}^s_c$  can be  replaced by $\mathbf{k}$ without confusion.

\subsection{The quasiparticle and charge currents carried by the wave packet}\label{Appendix:2_3}
The above derivation generalizes the derivation in \cite{SuNiu1999} to describe a Bogoliubov quasiparticle. As we have emphasized, the electric current carried by the Bogoliubov wave packet is quite different from the electric current carried by the  wave packet in a metal and this makes the problem quite subtle.
To proceed, we study the connection between the velocity operators and mass and charge positions. The position and charge position operators of the Bogoliubov quasiparticle  are defined as $\mathbf{r}I$ and $\mathbf{r}\tau^3$, respectively (see Appendix \ref{Appendix:1}). 
In the second quantized form,  
\begin{eqnarray}
&&\hat{\mathbf{r}}=\int\mathrm{d}\mathbf{r}~\mathbf{r}[c^\dagger_\uparrow(\mathbf{r})c_\uparrow(\mathbf{r})
+c_\downarrow(\mathbf{r})c^\dagger_\downarrow(\mathbf{r})],
\end{eqnarray}
and 
\begin{eqnarray}
&&\hat{\mathbf{r}}\tau^3=\int\mathrm{d}\mathbf{r}~\mathbf{r}[c^\dagger_\uparrow(\mathbf{r})c_\uparrow(\mathbf{r})
-c_\downarrow(\mathbf{r})c^\dagger_\downarrow(\mathbf{r})].
\end{eqnarray}
The BdG Hamiltonian Eq.~\eqref{Eq:BdG_Ham_Sec_G} can be separated into a noninteracting part and a pairing part, $H=H_{\mathrm{n}}+H_{\mathrm{p}}$. It is easy to check the commutation relations
\begin{eqnarray}
&&[\hat{\mathbf{r}},H_{\mathrm{n}}] =i\hat{\mathbf{v}}_\uparrow \tau^3,\\	
&&[\hat{\mathbf{r}}\tau^3,H_{\mathrm{n}}] =i\hat{\mathbf{v}}_\uparrow I ,\\	
&& [\hat{\mathbf{r}},H_{\mathrm{p}}] =\int\mathrm{d}\mathbf{r}\mathrm{d}\mathbf{r}'~
[(\mathbf{r-r'})\Delta(\mathbf{r},\mathbf{r}')\nonumber\\
&& ~~~~~~~~~~~~~~~~~~~~~~~~\times c^\dag_\uparrow(\mathbf{r})c^\dag_\downarrow(\mathbf{r}')-\mathrm{H.c.}],\\
&& [\hat{\mathbf{r}}\tau^3,H_{\mathrm{p}}]
=\int\mathrm{d}\mathbf{r}\mathrm{d}\mathbf{r}'~
[(\mathbf{r+r'})\Delta(\mathbf{r},\mathbf{r}')\nonumber\\
&& ~~~~~~~~~~~~~~~~~~~~~~~~\times c^\dag_\uparrow(\mathbf{r})c^\dag_\downarrow(\mathbf{r}')-\mathrm{H.c.}],\\
&&~~~~~~~~~~~=-i\frac{\mathrm{d}H_{\mathrm{p}}}{\mathrm{d }\mathbf{q}},
\end{eqnarray}
from which we get the Heisenberg equations 
\begin{eqnarray}
&&\frac{\mathrm{d} \hat{\mathbf{r}}}{\mathrm{d} t}=\hat{\mathbf{v}}_{qp}=-i[\hat{\mathbf{r}},H]=\hat{\mathbf{v}}_\uparrow \tau^3-i[\hat{\mathbf{r}},H_{\mathrm{p}}],\\ 
&&\frac{\mathrm{d} \hat{\mathbf{r}}\tau^3}{\mathrm{d} t}=-i[\hat{\mathbf{r}}\tau^3,H]=\hat{\mathbf{v}}_e-\frac{\mathrm{d}H_{\mathrm{p}}}{\mathrm{d }\mathbf{q}}.
\end{eqnarray} 
For a local pairing potential, $\Delta(\mathbf{r},\mathbf{r}')=\Delta_0(\mathbf{r})\delta_{[\mathbf{r}],[\mathbf{r}']}$, $\hat{\mathbf{r}}$ commutes with $H_{\mathrm{p}}$ and therefore the quasiparticle velocity operator $\hat{\mathbf{v}}_{qp}$ reduces to $\hat{\mathbf{v}}_\uparrow \tau^3$.

  The quasiparticle current is directly given by $\dot{\mathbf{r}}^s_c$,
  \begin{eqnarray}
   \mathbf{j}^s_{qp,\mathbf{k}}(\mathbf{q})&=&\dot{\mathbf{r}}^s_c=\partial_{\mathbf{k}} \mathcal{E}^s_{\mathbf{k}}(\mathbf{q}),\label{Eq:jqp}
  \end{eqnarray}
  which in the small $\mathbf{q}$ limit is
  \begin{eqnarray}
  j^s_{qp,\mathbf{k},i}&=&s\partial_{i}E_{\mathbf{k}}+\partial_{i}\partial_{j}\varepsilon_{\mathbf{k}}q_j,\label{Eq:jqp_2} \\
   &=&s\frac{\xi_\mathbf{k} \partial_{i}\varepsilon_\mathbf{k}}{E_\mathbf{k}}+s\frac{|\chi_\mathbf{k}|^2\Delta_{0}(\mathbf{k}) \partial_{i}\Delta_{0}(\mathbf{k})}{E_\mathbf{k}}\nonumber\\
   &&+s\frac{\Delta^2_{0}(\mathbf{k}) \partial_{i}|\chi_\mathbf{k}|^2}{2E_\mathbf{k}}+\partial_{i}\partial_{j}\varepsilon_{\mathbf{k}}q_j.\label{Eq:jqp_3}
  \end{eqnarray}
  Here we have used that $E_{\mathbf{k}}=\sqrt{(\varepsilon_{\mathbf{k}}-i)^2+\Delta^2_0(\mathbf{k})|\chi_\mathbf{k}|^2}$.
    The second term in Eq.~\eqref{Eq:jqp_3} is actually a multiband effect because
    \begin{eqnarray}
    \partial_i \Delta_{0}(\mathbf{k})&=&\partial_i\langle m_\mathbf{k}| \Delta(\mathbf{r})|m_\mathbf{k}\rangle,\\
    &=&\langle \partial_i m_\mathbf{k}| \Delta(\mathbf{r})|m_\mathbf{k}\rangle+\mathrm{H.c.},\\
    &=&\sum_{n\ne m}\langle \partial_i m_\mathbf{k}| n_\mathbf{k}\rangle\langle n_\mathbf{k}| \Delta(\mathbf{r})|m_\mathbf{k}\rangle+\mathrm{H.c.}.~~
    \end{eqnarray}
    Note that $\langle n_\mathbf{k}| \Delta(\mathbf{r})|m_\mathbf{k}\rangle$ is the interband pairing, which vanishes for position independent pairing potential $\Delta_0(\mathbf{r})=\Delta_0$ and $\langle \partial_i m_\mathbf{k}| n_\mathbf{k}\rangle$ is proportional to the interband matrix element of the single particle velocity operator \cite{PhysRevB.95.024515}, 
    \begin{eqnarray}
        &&\int\mathrm{d}\mathbf{r}~  m^\ast_\mathbf{k}(\mathbf{r})e^{-i\mathbf{k}\cdot\mathbf{r}}\hat{v}_{\uparrow,i} e^{i\mathbf{k}\cdot\mathbf{r}}n_\mathbf{k}(\mathbf{r})\nonumber\\
        &&=\int_{\mathrm{u.c.}}\mathrm{d}\mathbf{r}~  m^\ast_\mathbf{k}(\mathbf{r})\partial_{i}h_{\mathbf{k}}(\mathbf{r})n_\mathbf{k}(\mathbf{r}),\nonumber\\
        &&=\partial_i \varepsilon_{m\mathbf{k}} \delta_{mn}+(\varepsilon_{m\mathbf{k}}-\varepsilon_{n\mathbf{k}})\langle \partial_i m_\mathbf{k}| n_\mathbf{k}\rangle,
    \end{eqnarray}
where $h_{\mathbf{k}}(\mathbf{r})=e^{-i\mathbf{k}\cdot\mathbf{r}}h(\mathbf{r}) e^{i\mathbf{k}\cdot\mathbf{r}}$ is the Bloch Hamiltonian.
  
     On the other hand, when the pairing potential is nonlocal, we obtain
     \begin{eqnarray}
      \mathbf{j'}^s_{qp,\mathbf{k}}(\mathbf{q})&=&\dot{\mathbf{r}}^s_c+i\langle \Psi^s_{\mathbf{k}}|[\hat{\mathbf{r}},H_{\mathrm{p}}]| \Psi^s_{\mathbf{k}}\rangle,\\
      &=&\partial_{\mathbf{k}} \mathcal{E}^s_{\mathbf{k}}(\mathbf{q})-[u^s_{\mathbf{k}}v^s_{\mathbf{k}}\Delta_{0,\mathbf{k}}(\mathbf{q})\partial_{\mathbf{k}}\chi_\mathbf{k}+\mathrm{H.c.}],~~~\\
      &=&\partial_{\mathbf{k}} \mathcal{E}^s_{\mathbf{k}}(\mathbf{q})-s\frac{|\Delta_{0,\mathbf{k}}(\mathbf{q})|^2 \partial_{\mathbf{k}}|\chi_\mathbf{k}|^2}{2E_\mathbf{k}(\mathbf{q})}.\label{Eq:jqpp}
     \end{eqnarray}
 Clearly, $\sum_{s,\mathbf{k}}\mathbf{j}^s_{qp,\mathbf{k}}(\mathbf{q})=\sum_{s,\mathbf{k}}\mathbf{j'}^s_{qp,\mathbf{k}}(\mathbf{q})$, see Eqs.~\eqref{Eq:jqp} and ~\eqref{Eq:jqpp} and consider summation over $s$. Therefore $\mathbf{j'}_{qp,\alpha}$ can be replaced by $\mathbf{j}_{qp,\alpha}$ in Eq.~\eqref{Eq:j_t_a}.
   
The quasiparticle charge current can be calculated as
\begin{eqnarray}
\mathbf{j}^s_{e,\mathbf{k}}(\mathbf{q})=\dot{\mathbf{r}}^s_e+\langle \Psi^s_{\mathbf{k}}|\frac{\mathrm{d}H_{\mathrm{p}}}{\mathrm{d}\mathbf{q}}| \Psi^s_{\mathbf{k}}\rangle.
\end{eqnarray}
The first term in the above equation is
\begin{eqnarray}
\dot{\mathbf{r}}^s_e=(|u^s_{\mathbf{k}}|^2-|v^s_{\mathbf{k}}|^2)\dot{\mathbf{r}}^s_c, \label{Eq:v_e_1}
\end{eqnarray} 
 and the second term can be calculated using the relation
\begin{eqnarray}
\langle \Psi^s_{\mathbf{k}}|\frac{\mathrm{d}H_{\mathrm{p}}}{\mathrm{d}\mathbf{q}}| \Psi^s_{\mathbf{k}}\rangle&=&\frac{\mathrm{d}}{\mathrm{d}\mathbf{q}}\langle \Psi^s_{\mathbf{k}}|H_{\mathrm{p}}| \Psi^s_{\mathbf{k}}\rangle\label{Eq:v_a}\\
&&-\bigg(\langle \frac{\mathrm{d}}{\mathrm{d}\mathbf{q}}\Psi^s_{\mathbf{k}}|H_{\mathrm{p}}| \Psi^s_{\mathbf{k}}\rangle+\mathrm{H.c.}\bigg).~\label{Eq:v_e_2}
\end{eqnarray}
After some calculations, Eq.~\eqref{Eq:v_a} gives the anomalous velocity
\begin{eqnarray}
v^s_{a,\mathbf{k},i}&=&2s|\Delta_{\mathbf{k}}u^s_{\mathbf{k}}v^s_{\mathbf{k}}|\frac{\mathrm{d}}{\mathrm{d}q_i}e^{-\bar{g}_{jl}(\mathbf{k})q_j q_l},\\
&\approx&-2s\frac{|\Delta_{\mathbf{k}}|^2}{E_{\mathbf{k}}}\bar{g}_{ij}q_{j},\label{Eq:v_a_2}
\end{eqnarray}
and  Eq.~\eqref{Eq:v_e_1} and Eq.~\eqref{Eq:v_e_2} give the conventional velocity,
\begin{eqnarray}
v^s_{e,\mathbf{k},i}&=&(|u^s_{\mathbf{k}}|^2-|v^s_{\mathbf{k}}|^2) \dot{r}^s_{c,i}\\
&&-2s[e^{-\bar{g}_{jl}q_j q_l}|u^s_{\mathbf{k}}| \frac{\mathrm{d}|v^s_{\mathbf{k}}|}{\mathrm{d}k_{i}}-(u^s\leftrightarrow v^s) ],\nonumber\\
&\approx&\partial_i \varepsilon_\mathbf{k}+s \frac{\xi_{\mathbf{k}}}{E_{\mathbf{k}}}\partial_i\partial_j \varepsilon_\mathbf{k} q_j.\label{Eq:v_e}
\end{eqnarray}

There is a simpler way to obtain the same result in a more intuitive form. Noting that the noninteracting part of the Hamiltonian is $\mathbf{q}$ independent,  we have
\begin{eqnarray}
\frac{\mathrm{d}H_{\mathrm{p}}}{\mathrm{d}\mathbf{q}}=\frac{\mathrm{d}H}{\mathrm{d}\mathbf{q}},
\end{eqnarray} 
 and then
\begin{eqnarray}
\mathbf{j}^s_{e,\mathbf{k}}(\mathbf{q})&=&\dot{\mathbf{r}}^s_e+\langle \Psi^s_{\mathbf{k}}|\frac{\mathrm{d}H}{\mathrm{d}\mathbf{q}}| \Psi^s_{\mathbf{k}}\rangle,\\
&=&\frac{\mathrm{d}}{\mathrm{d}\mathbf{q}}\langle \Psi^s_{\mathbf{k}}|H| \Psi^s_{\mathbf{k}}\rangle\\
&&-i\langle \Psi^s_{\mathbf{k}}|[\hat{\mathbf{r}}\tau^3,H]| \Psi^s_{\mathbf{k}}\rangle-\bigg(\langle \Psi^s_{\mathbf{k}}|H| \frac{\mathrm{d}}{\mathrm{d}\mathbf{q}}\Psi^s\rangle+\mathrm{H.c.}\bigg).\nonumber~~\label{Eq:zero} 
\end{eqnarray}
Substituting  
 \begin{eqnarray}
  \frac{\mathrm{d}\Psi^s_{\mathbf{k}}(\mathbf{r})}{\mathrm{d}\mathbf{q}}&=& i\mathbf{r}\tau^3\Psi^s_{\mathbf{k}}(\mathbf{r})\\
  &&+\int \mathrm{d}\mathbf{p}~W^s_\mathbf{p}e^{i\mathbf{p}\cdot \mathbf{r}}\left[\begin{array}{c}
  u_\mathbf{p}e^{i \mathbf{q} \cdot\mathbf{r}}\frac{\mathrm{d}m_\mathbf{p+q}(\mathbf{r})}{\mathrm{d}\mathbf{q}} \\ 
  v_\mathbf{p}e^{-i \mathbf{q} \cdot\mathbf{r}}\frac{\mathrm{d}m_\mathbf{p-q}(\mathbf{r})}{\mathrm{d}\mathbf{q}}
  \end{array} \right],\nonumber\label{Eq:dPsi}
 \end{eqnarray} 
into Eq.~\eqref{Eq:zero}, we find
\begin{widetext}
 \begin{eqnarray}
 &&-i\langle \Psi^s_{\mathbf{k}}|[\hat{\mathbf{r}}\tau^3,H]| \Psi^s_{\mathbf{k}}\rangle
  -\bigg(\langle \Psi^s_{\mathbf{k}}|H| \frac{\mathrm{d}}{\mathrm{d}\mathbf{q}}\Psi^s\rangle+\mathrm{H.c.}\bigg)\nonumber\\
  &=&-i\int \mathrm{d}\mathbf{r}~\mathbf{r}\Psi^{s\ast}_{\mathbf{k}}(\mathbf{r})(\tau^3H-H\tau^3)\Psi^{s}_{\mathbf{k}}(\mathbf{r})-i\int \mathrm{d}\mathbf{r}~\mathbf{r}\Psi^{s\ast}_{\mathbf{k}}(\mathbf{r})(H\tau^3-\tau^3H)\Psi^{s}_{\mathbf{k}}(\mathbf{r})\nonumber\\
 &&-\int\mathrm{d}\mathbf{r}\mathrm{d}\mathbf{p}\mathrm{d}\mathbf{p}'~W^s_\mathbf{p}W^{s\ast}_\mathbf{p'}\mathcal{E}^s_{\mathbf{p}'}e^{i\mathbf{(p-p')\cdot r}}\bigg[u_\mathbf{p}u^\ast_{\mathbf{p}'} m^\ast_\mathbf{p'+q}(\mathbf{r})\frac{\mathrm{d}m_\mathbf{p+q}(\mathbf{r})}{\mathrm{d}\mathbf{q}} + 
 v_\mathbf{p}v^\ast_{\mathbf{p}'} m^\ast_\mathbf{p'-q}(\mathbf{r})\frac{\mathrm{d}m_\mathbf{p-q}(\mathbf{r})}{\mathrm{d}\mathbf{q}}\bigg]+\mathrm{H.c.},\\
 &=&-\int\mathrm{d}\mathbf{p}~W^s_\mathbf{p}W^{s\ast}_\mathbf{p}\mathcal{E}^s_{\mathbf{p}}\bigg[u_\mathbf{p}u^\ast_{\mathbf{p}}\langle m_\mathbf{p+q}|\frac{\mathrm{d}}{\mathrm{d}\mathbf{q}}|m_\mathbf{p+q}\rangle + 
 v_\mathbf{p}v^\ast_{\mathbf{p}} \langle m_\mathbf{p-q}|\frac{\mathrm{d}}{\mathrm{d}\mathbf{q}}|m_\mathbf{p-q}\rangle+\mathrm{H.c.}\bigg]=0.
 \end{eqnarray}
 \end{widetext}
 Therefore we obtain that 
 \begin{eqnarray}
 \mathbf{j}^s_{e,\mathbf{k}}(\mathbf{q}) 
 =\frac{\mathrm{d}}{\mathrm{d}\mathbf{q}}\langle \Psi^s_{\mathbf{k}}|H| \Psi^s_{\mathbf{k}}\rangle=\frac{\mathrm{d}\mathcal{E}^s_{\mathbf{k}}(\mathbf{q})}{\mathrm{d}\mathbf{q}}.\label{Eq:je_2}
 \end{eqnarray}
As we mentioned, $\mathbf{q}$ is the momentum of the Cooper pair, and $\mathbf{k}$ is the momentum of Bogoliubov quasiparticle, and therefore $\mathcal{E}^s_{\mathbf{k}}(\mathbf{q})$ can be viewed as the dispersion of both the quasiparticle and the Cooper pair. The quasiparticle current is given by the group velocity of the quasiparticle, Eq.~\eqref{Eq:jqp}, while the charge current is given by the group velocity of the Cooper pair, Eq.~\eqref{Eq:je_2}.

Superficially, one may think that the wave packet   can be replaced by $|\psi_\mathbf{k}\rangle$ in the above calculations. However, evaluating the position operator on the Bloch-like state $|\psi_\mathbf{k}\rangle$ gives an ill-defined result \cite{PhysRevLett.80.1800}, and as we will show, the anomalous velocity is absent when evaluating the operator $\hat{\mathbf{v}}_e$ on $|\psi_\mathbf{k}\rangle$ directly. Therefore, the wave packet with a well defined position is needed, at least conceptually.

As a direct application of our results, we study the superfluid weight. The total electric current in the small $\mathbf{q}$ limit is
\begin{eqnarray}
j_i&=&-\frac{1}{2}\sum_{s,\mathbf{k}}\tanh{\bigg(\frac{\beta \mathcal{E}^s_\mathbf{k}}{2}\bigg)}j^s_{e,\mathbf{k},i}+\frac{1}{2}\sum_{s,\mathbf{k}}j^s_{qp,\mathbf{k},i},\\
&\approx&\sum_{\mathbf{k}}\bigg[\partial_{i}\partial_{j}\varepsilon_{\mathbf{k}}-\frac{\xi_{\mathbf{k}}\tanh{(\beta E_\mathbf{k}/2)}}{E_\mathbf{k}}\partial_{i}\partial_{j}\varepsilon_{\mathbf{k}}\nonumber\\
&&~~~~~~~~~-\frac{\beta\partial_i\varepsilon_\mathbf{k}\partial_j\varepsilon_\mathbf{k}}{2\cosh^2{(\beta E_\mathbf{k}/2)}}\bigg]q_j\\
&&+2
\sum_{\mathbf{k}}\frac{|\Delta_{\mathbf{k}}|^2\tanh{(\beta E_\mathbf{k}/2)}}{E_\mathbf{k}}\bar{g}_{ij}(\mathbf{k})q_j.
\end{eqnarray}
 The coefficient relating $j_i$ and $q_j$ gives the superfluid weight, which can be separated into conventional and geometric parts \cite{PT,LiebLattice,PhysRevB.95.024515}
 \begin{eqnarray}
 D_{ij}=D_{\mathrm{conv},ij}+D_{\mathrm{geom},ij},
 \end{eqnarray}
with 
\begin{eqnarray}
D_{\mathrm{conv},ij}&=&\sum_{\mathbf{k}}\bigg[\partial_{i}\partial_{j}\varepsilon_{\mathbf{k}}-\frac{\xi_{\mathbf{k}}\tanh{(\beta E_\mathbf{k}/2)}}{E_\mathbf{k}}\partial_{i}\partial_{j}\varepsilon_{\mathbf{k}}\nonumber\\
&&~~~~~~~~~-\frac{\beta\partial_i\varepsilon_\mathbf{k}\partial_j\varepsilon_\mathbf{k}}{2\cosh^2{(\beta E_\mathbf{k}/2)}}\bigg],\label{Eq:SW_conv}
\end{eqnarray}
and 
\begin{eqnarray}
D_{\mathrm{geom},ij}&=&2
\sum_{\mathbf{k}}\frac{|\Delta_{\mathbf{k}}|^2\tanh{(\beta E_\mathbf{k}/2)}}{E_\mathbf{k}}\bar{g}_{ij}(\mathbf{k}).
\end{eqnarray}
 The geometric term obtained in this paper is a generalization of previous results \cite{PT,LiebLattice,PhysRevB.95.024515}, where the pairing potential  was restricted to be $\Delta(\mathbf{r},\mathbf{r}')=\Delta_0\delta_{\mathbf{r},\mathbf{r}'}$. For continuum systems without periodic potentials, $m_{\mathbf{k}}(\mathbf{r})$ is a constant, and therefore $\bar{g}_{ij}(\mathbf{k})$ vanishes and the geometric term is absent.
The first term 
in Eq.~\eqref{Eq:SW_conv} stems from the quasiparticle current $\mathbf{j}_{qp}$. It is zero in the presence of the periodic potential. 
In the continuum limit, $\varepsilon_{\mathbf{k}}=\mathbf{k}^2/(2m)$,  here $m$ is the mass of the particle. Then for $i=j$, $\sum_{\mathbf{k}}\partial_{i}\partial_{j}\varepsilon_{\mathbf{k}}$ diverges  and cancels the divergence in the second term in Eq.~\eqref{Eq:SW_conv}. 
\begin{eqnarray}
D_{ij}&=&\frac{\delta_{ij}}{m}\int\frac{\mathrm{d}\mathbf{k}}{(2\pi)^d}\bigg[1-\frac{\xi_{\mathbf{k}}\tanh{(\beta E_\mathbf{k}/2)}}{E_\mathbf{k}}\nonumber\\
&&~~~~~~~~~-\frac{1}{d}\frac{\beta \varepsilon_{\mathbf{k}}}{\cosh^2{(\beta E_\mathbf{k}/2)}}\bigg],\\
&=&\frac{\delta_{ij}}{m}\bigg[n-\frac{1}{d}\int\frac{\mathrm{d}\mathbf{k}}{(2\pi)^d}\frac{\beta \varepsilon_{\mathbf{k}}}{\cosh^2{(\beta E_\mathbf{k}/2)}}\bigg],
\end{eqnarray}
where $n$ is the particle density and $d$ is the spatial dimension of the system. This recovers the well-known mean-field result for the superfluid weight in the continuum limit \cite{PhysRev.108.1175}. 

 \subsection{Comparison to the fully quantum mechanical derivation}\label{Appendix:2_4}
 Using the semiclassical wave packet approach we have shown that  the quasiparticle and charge currents are given by the group velocities of the quasiparticle and the Cooper pair, respectively. 
 Since $\mathcal{E}^s_{\mathbf{k}}(\mathbf{q})$ is the  energy corresponding to the wave function, Eq.~\eqref{Eq:qp_wavefunction_a}, 
one may think that the same results can be obtained by evaluating  the currents  $\mathbf{j}_{e}$  and $\mathbf{j}_{qp}$  using the wave function Eq.~\eqref{Eq:qp_wavefunction_a}. However, direct calculations show 
\begin{eqnarray}
\mathbf{j}^s_{e,\mathbf{k}} &=&\langle \psi^s_{\mathbf{k}}| \hat{\mathbf{v}}_e|\psi^s_{\mathbf{k}}\rangle ,\nonumber\\
&=&|u^s_{\mathbf{k}}|^2\partial_{\mathbf{k}}\varepsilon_{\mathbf{k+q}}+|v^s_{\mathbf{k}}|^2\partial_{\mathbf{k}}\varepsilon_{\mathbf{k-q}},
\end{eqnarray}
and 
\begin{eqnarray}
\mathbf{j}^s_{qp,\mathbf{k}}&=&\langle \psi^s_{\mathbf{k}}| \hat{\mathbf{v}}_\uparrow \tau^3-i[\hat{\mathbf{r}},H_{\mathrm{p}}]|\psi^s_{\mathbf{k}}\rangle,\nonumber\\
&=&|u^s_{\mathbf{k}}|^2\partial_{\mathbf{k}}\varepsilon_{\mathbf{k+q}}-|v^s_{\mathbf{k}}|^2\partial_{\mathbf{k}}\varepsilon_{\mathbf{k-q}}\nonumber\\
&&+s\frac{|\Delta_{0,\mathbf{k}}(\mathbf{q})|^2 \partial_{\mathbf{k}}|\chi_\mathbf{k}|^2}{2E_\mathbf{k}(\mathbf{q})}.
\end{eqnarray}

In the small $\mathbf{q}$ limit, we find 
\begin{eqnarray}
j^s_{qp,\mathbf{k},i}
&=&s\frac{\xi_{\mathbf{k}}\partial_{i}\varepsilon_\mathbf{k}}{E_{\mathbf{k}}}+s\frac{\Delta^2_{0}(\mathbf{k}) \partial_{i}|\chi_\mathbf{k}|^2}{2E_\mathbf{k}}\nonumber\\
&&+\partial_{i}\partial_{j}\varepsilon_{\mathbf{k}}q_j,~~~~\label{Eq:jqp_wrong}
\end{eqnarray}
and 
\begin{eqnarray}
&&j^s_{e,\mathbf{k},i}=\partial_i \varepsilon_\mathbf{k}+s \frac{\xi_{\mathbf{k}}}{E_{\mathbf{k}}}\partial_i\partial_j \varepsilon_\mathbf{k} q_j.\label{Eq:je_wrong} 
\end{eqnarray}
Comparing to Eqs.~\eqref{Eq:jqp_3}, ~\eqref{Eq:v_a_2} and~\eqref{Eq:v_e}, we see that Eq.~\eqref{Eq:jqp_wrong} is correct only for momentum independent $\Delta_0(\mathbf{k})$ and 
the anomalous velocity is missing in Eq.~\eqref{Eq:je_wrong}. The reason is that the isolated band wave function Eq.~\eqref{Eq:qp_wavefunction_a} is accurate only up to the zeroth order of the inverse band gap and the interband processes  are not taken into account. For position dependent $\Delta_0(\mathbf{r})$, in general there will be interband pairing, $\Delta_{0,mn}(\mathbf{k})=\langle m_\mathbf{k}|\Delta_0(\mathbf{r})|n_\mathbf{k}\rangle$, which gives corrections to the wave function Eq.~\eqref{Eq:qp_wavefunction_a} even in the $\mathbf{q}=0$ limit and leads to the second term in Eq.~\eqref{Eq:jqp_3}. More importantly, a nonzero phase twist  also induces interband pairings and gives rise to the quantum metric correction to the charge current in the isolated band limit \cite{PhysRevB.95.024515}. 
To get the correct result in the fully quantum mechanical approach, we have to solve the BdG equation {\it by including all the bands and take the isolated band limit after obtaining the currents.} 
The physics behind this  procedure is opaque and for general multiband systems with nonuniform pairing potentials, this approach  is difficult to apply. 
 On the other hand, the (lowest order) multiband effects have been incorporated in the energy $\mathcal{E}^s_{\mathbf{k}}(\mathbf{q})$, because the first order correction to the energy is obtained using the zeroth order wave function. Using the semiclassical approach the currents are expressed in terms of $\mathcal{E}^s_{\mathbf{k}}(\mathbf{q})$, and therefore the multiband effects appear naturally.

\section{Mean-field theory for the attractive Hubbard model on the sawtooth lattice}\label{Appendix:3}

\begin{figure}
	\includegraphics[width=0.9\columnwidth]{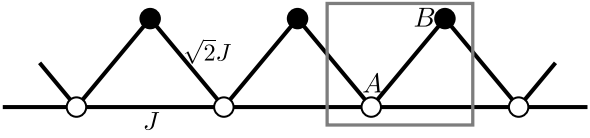}
	\caption{Sawtooth lattice and its unit cell (grey box). The orbitals in the unit cell are labeled by $\alpha=A, B$. 
	}\label{Fig:Saw_Tooth_l}
\end{figure}

The attractive Hubbard model on the sawtooth lattice is defined through the Hamiltonian
\begin{eqnarray}
 H=H_{\mathrm{kin}}-\mu N+H_{\mathrm{int}}.
\end{eqnarray}
 Where the noninteracting term is 
 \begin{eqnarray}
 H_{\mathrm{kin}}-\mu N=\sum_{k,\sigma}\mathbf{c}^\dag_{k\sigma} h_0(k)\mathbf{c}_{k\sigma},
 \end{eqnarray}
  with the hopping matrix given by (see Fig.~\ref{Fig:Saw_Tooth_l})
\begin{eqnarray}
h_0(k)=\left[
\begin{array}{cc}
2J\cos{k}-\mu & 2\sqrt{2}J\cos{\frac{k}{2}} \\
2J\sqrt{2}\cos{\frac{k}{2}} & -\mu 
\end{array}
\right].
\end{eqnarray}
 The operators are defined as $\mathbf{c}^{\dag}_{k\sigma}=(c^{\dag}_{Ak\sigma},c^{\dag}_{Bk\sigma})$, and
 \begin{eqnarray}
 c^{\dag}_{\alpha k\sigma}=\frac{1}{\sqrt{N_c}}\sum_i e^{i k r_{i\alpha}}c^{\dag}_{i\alpha\sigma},
 \end{eqnarray}
 where $N_c$ is the number of unit cells, $r_{i\alpha}$ is the position of the $\alpha$ orbital in the $i$-th unit cell and $c^{\dag}_{i\alpha\sigma}$ creates a fermion with spin $\sigma=\uparrow,\downarrow$ at $r_{i\alpha}$. Solving the eigenvalue problem, we get the band dispersions 
\begin{eqnarray}
 &&\xi_{-,k}=-2J-\mu,~~
\xi_{+,k}=2J(1+\cos{k})-\mu.
\end{eqnarray}
 The quantum metrics of the two bands are the same 
 \begin{eqnarray}
g=\frac{1-\cos{k}}{2(2+\cos{k})^2}.
 \end{eqnarray}
 The attractive Hubbard interaction
 \begin{eqnarray}
  H_{\mathrm{int}}=-U\sum_{i\alpha}n_{i\alpha\uparrow}n_{i\alpha\downarrow},
 \end{eqnarray} 
  with $U>0$,  can be approximated by  
\begin{eqnarray}
H_{\mathrm{int}}\approx \sum_{i\alpha}(\Delta_{\alpha}c^\dag_{i\alpha\uparrow}c^\dag_{i\alpha\downarrow}+\mathrm{H.c.})+U\sum_{i\alpha\sigma}n_\alpha n_{i\alpha\sigma},
\end{eqnarray}
with the pairing potential $\Delta_{\alpha}=-U\langle c_{i\alpha\downarrow}c_{i\alpha\uparrow} \rangle$ and the Hartree potential $U n_\alpha=U \langle n_{i\alpha\sigma}\rangle$. The inequivalence of $A$ and $B$ indicates that the order parameters  on the two orbitals are different.  

\begin{figure}
	\includegraphics[width=1\columnwidth]{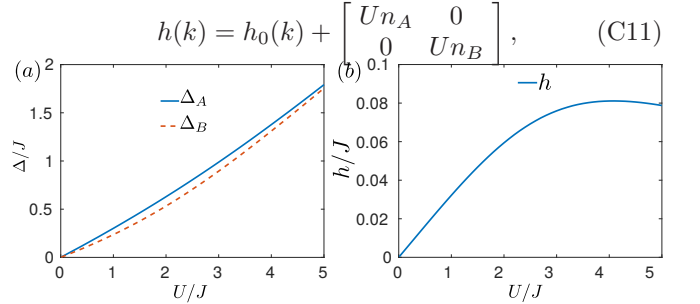}
	\caption{Pairing potentials (a) and the difference of the Hartree fields (b) as functions of $U$. The filling is chosen such that the flat band is half-filled in the noninteracting limit. 
	}\label{Fig:Saw_Tooth_order}
\end{figure} 

Within the mean-field approximation, we get the BdG Hamiltonian
\begin{eqnarray}
H=\sum_{k}\mathbf{C}^\dag_{k} \mathcal{H}_k \mathbf{C}_{k},
\end{eqnarray}
  with $\mathbf{C}^{\dag}_{k}=[\mathbf{c}^{\dag}_{k\uparrow},(\mathbf{c}_{-k\downarrow})^T]$ and
\begin{eqnarray}
\mathcal{H}_k=\left[
\begin{array}{cc}
h(k) & \Delta \\
\Delta & -h(k)
\end{array}
\right],
\end{eqnarray}
where $\Delta=diag(\Delta_A,\Delta_B)$ and 
\begin{eqnarray}
h(k)=h_0(k)+\left[
\begin{array}{cc}
Un_A & 0 \\
0 & Un_B
\end{array}
\right],
\end{eqnarray}
the dispersions in the presence of the Hartree field become
\begin{eqnarray}
\xi_{\pm, k}=&&J\cos{k}-\mu_{\mathrm{eff}}
\pm \sqrt{[J(\cos{k}+2)+h]^2-4Jh},\nonumber\\
\end{eqnarray}
with $\mu_{\mathrm{eff}}=\mu-U(n_A+n_B)/2$ and $h=U(n_A-n_B)/2$. 
The parameters $\Delta_{\alpha}$ and $n_{\alpha}$ should be determined self-consistently. Fig.~\ref{Fig:Saw_Tooth_order} shows $\Delta_A$, $\Delta_B$ and $h$ as functions of $U$. The filling is chosen such that the flat band is half-filled in the noninteracting limit. The pairing potentials increase linearly with increasing $U$, while the Hartree field difference $h$ has a nonmonotonic  behavior, due to the interplay between the  kinetic energy and Hubbard interaction.

\bibliography{paper_draft.bib}
\end{document}